\documentclass[%
 reprint,
 amsmath,amssymb,
 aps,
 pre,
floatfix,
]{revtex4-2}

\usepackage{graphicx}
\usepackage{xcolor}
\usepackage{amsmath}
\usepackage{dcolumn}
\usepackage{bm}
\usepackage{subcaption}
\usepackage{url}

\begin{document}

\title{Competing Social Contagions with Opinion Dependent Infectivity}

\author{Corbit R. Sampson}
 \affiliation{Department of Applied Mathematics, University of Colorado at Boulder, Colorado 80309, USA}
 
 \author{Juan G. Restrepo}
 \affiliation{Department of Applied Mathematics, University of Colorado at Boulder, Colorado 80309, USA}

\date{\today}

\begin{abstract}
The spread of disinformation (maliciously spread false information) in online social networks has become an important problem in today's society. Disinformation's spread is facilitated by the fact that individuals often accept false information based on cognitive biases which predispose them to believe information that they have heard repeatedly or that aligns with their beliefs. Moreover, disinformation often spreads in direct competition with a corresponding true information. To model these phenomena, we develop a model for two competing beliefs spreading on a social network, where individuals have an internal opinion that models their cognitive biases and modulates their likelihood of adopting one of the competing beliefs. By numerical simulations of an agent-based model and a mean-field description of the dynamics, we study how the long-term dynamics of the spreading process depends on the initial conditions for the number of spreaders and the initial opinion of the population. We find that the addition of cognitive biases enriches the transient dynamics of the spreading process, facilitating behavior such as the revival of a dying belief and the overturning of an initially widespread opinion. Finally, we study how external recruitment of spreaders can lead to the eventual dominance of one of the two beliefs.
\end{abstract}

\maketitle

\section{\label{introduction}Introduction}
In the last few decades, social media has become increasingly more ubiquitous in people's lives \cite{edosomwan2011history,sajithra2013social}. Online social media has become a source of news for many individuals, with about half of US adults admitting to receiving news at least ``occasionally" through social media \cite{shearer2018news, chadwick2019news, PewResearch2023}.  However, the widespread use of social media and other factors such as its low barrier to entry, limited view format, and ideologically segregated social networks makes online social media platforms an attractive target for the malicious dissemination of false information  (known as disinformation)  \cite{allcott2017social,bakshy2015exposure,di2021fake}. The spread of disinformation and misinformation (the unintentional spread of false or inaccurate information) has been labeled a major threat to national security and appears as a concern relating to health security, political instability, and violent societal conflict \cite{blair2010annual}. These concerns have led to great interest in the study of how disinformation and misinformation spread  \cite{gelfert2018fake,berthon2018brands,rini2017fake,allcott2017social,lazer2018science,pennycook2018prior,bronstein2019belief,mcdermott2019psychological,allcott2017social,vicario2019polarization,tandoc2018audiences,berthon2018brands,cheng2021influence,chen2020consumer,vafeiadis2020refuting, franceschi2022spreading,murayama2021modeling} and in the development of methods to limit the spread of misinformation \cite{rabb2022cognitive,tambuscio2015fact,budak2011limiting,zareie2021minimizing,van2022misinformation,cook2015misinformation}.

Although the spread of disinformation within online social media platforms can be exacerbated by many mechanisms, here we are interested in understanding the effect of individuals' cognitive biases in the spread of disinformation. A cognitive bias is the tendency for human cognition to consistently form beliefs that are systematically distorted from reality \cite{bias2015evolution}. Particularly, we are interested in the effects of the {\it confirmation bias} and the {\it illusory truth effect}. Confirmation bias is the tendency of individuals to more readily believe information which aligns better with their own beliefs \cite{hart2009feeling}. The illusory truth effect is the tendency of individuals to view ideas as more truthful through mere exposure (i.e., exposure to those ideas without additional reinforcement) \cite{hasher1977frequency, pennycook2018prior}. Together, these two biases lead to the possibility that individuals may  believe a particular piece of information simply from repeated exposure. The effects of repeated exposure promoting the spread of misinformation have been observed in relation to public health misinformation involving COVID-19 \cite{weng2024repeated}.

In this study we develop an agent-based model to examine how confirmation bias and the illusory truth effect can affect the spreading dynamics of two mutually exclusive beliefs, leading to the predominance of one over the other. 
In our model, the two competing beliefs are represented as two discrete states, $+1$ and $-1$. Adopting terminology from the social contagion and epidemic spreading literature, we refer to these states as {\it contagions}, and the adoption of one of these beliefs as an {\it infection}. In order to model confirmation bias and the illusory truth effect, each individual is endowed with an internal, continuous opinion variable, which represents the alignment of the individual's biases towards the competing beliefs. This internal opinion is modified by infection attempts, modeling the illusory truth effect, and modifies the infection probabilities, modeling confirmation bias. We study the long-term dynamics of the competing beliefs by means of numerical simulations of the agent-based model and a mean-field description of the dynamics. We find that there is a continuum of disease-free states, each characterized by a different average internal opinion of the population. The average internal opinion determines the stability of the disease-free state. As opposed to traditional spreading processes, the presence of cognitive biases can lead to unexpected dynamics depending on the initial conditions. In some situations, a pair of competing beliefs with numbers of supporters that are initially decaying can rebound, so that one of the beliefs ends up becoming dominant while the other dies. Similarly, a population with an average opinion that initially is biased towards one belief can end up overturning this opinion so that the opposing belief becomes dominant. We also study how the long-term dynamics is modified by external recruitment of spreaders for one of the two beliefs and find that, depending on the initial conditions, this recruitment can lead either to total domination by the promoted belief or coexistence of the two beliefs. 

There are a number of other studies that have examined the effects of multiple interacting contagions both in the context of biological and social contagions \cite{zhuang2017modeling,myers2012clash,fu2017dueling,kaligotla2015}. These have included both competitive and cooperative interactions \cite{zhuang2017modeling,myers2012clash} as well as the simultaneous spread of viral contagions and vaccination seeking behavior \cite{fu2017dueling}. Some studies have even included many heterogeneous features, such as the work by Kaligotla et al. \cite{kaligotla2015}, which developed a
threshold-like agent-based model of two competing rumors which included agent reputation, effort of information spreading, and contrarian agents. Although these previous studies examined multiple spreading contagions and their potentially complex interactions, in our study we also highlight the role of individual opinions and cognitive biases in the spread of competing beliefs.

Our paper proceeds as follows. In Sec.~\ref{modelanalysis}, we introduce our agent-based model. In Sec.~\ref{MFA}, we formulate a mean-field approximation of our model. In Sec.~\ref{STB}, we discuss the possible long-term behaviors of the model and classify their linear stability. In Sec.~\ref{behavior}, we discuss two of the primary behaviors of the model and how they arise from opinion dependent stability of the disease-free state. In Sec.~\ref{Opinion Distrobution} we study the effect of a heterogeneous distribution of initial opinions. In Sec.~\ref{reintroduction}, we modify the model to include external recruitment of spreaders. Finally, in Sec.~\ref{discussion}, we summarize and discuss our findings. The code for this project is available at \url{https://github.com/CorbitSampson/Competing_Social_Contagions}.  

\section{\label{modelanalysis}Description of the model}

We consider a model where individuals can adopt one of two mutually exclusive beliefs or remain neutral, and individuals who have adopted one of the two beliefs try to actively spread their belief to the rest of the population. In order to make contact with existing literature and terminology on social contagion and epidemic processes, we will refer to the two beliefs as ``contagions'', and label them as $+1$ and $-1$. We will refer to the neutral state as the ``susceptible'' state, and label it with a $0$.  Therefore, each individual has a trinary contagion state, either $-1$,$0$, or $+1$. We will also say that an individual who adopts one of the two beliefs is ``infected''. In addition to the contagion state, each node has an internal opinion which is used to model the effects of confirmation bias and the illusory truth effect. Confirmation bias is the effect whereby a person is more likely to believe information that already aligns with their current belief \cite{hart2009feeling}, and the illusory truth effect is a phenomenon where people are more likely to believe something if they have been repeatedly exposed to it \cite{hasher1977frequency, pennycook2018prior}. Below we describe our model in detail.

Our model consists of a network where, at time $t$, each node $i$ has a discrete contagion state $s^t_i\in \{-1,0,1 \}$. The contagion states  $-1$ and $+1$ indicate that the individual is infected with one of the two mutually exclusive contagions and can spread this contagion to its network neighbors. The state $0$ indicates the individual is susceptible. In addition, each node has a continuous internal opinion $x_i^t\in [-1,1]$. The opinion $x^t_i$ measures the node's alignment with each of the two contagions.  To model the effects of confirmation bias we will assume that, the closer $x_i^t$ is to $+1$ ($-1$), the more likely it is that node $i$ is infected with opinion $+1$ ($-1$) when exposed by a neighbor and the less likely it is to recover from it. Furthermore, to model the illusory truth effect, each time a node with contagion $s_i$ attempts to infect a node $j$, the opinion of node $j$ moves closer to $s_i$, even if the infection attempt is unsuccessful.
\begin{figure}[t]
    \centering
    \includegraphics[width=0.47\textwidth]{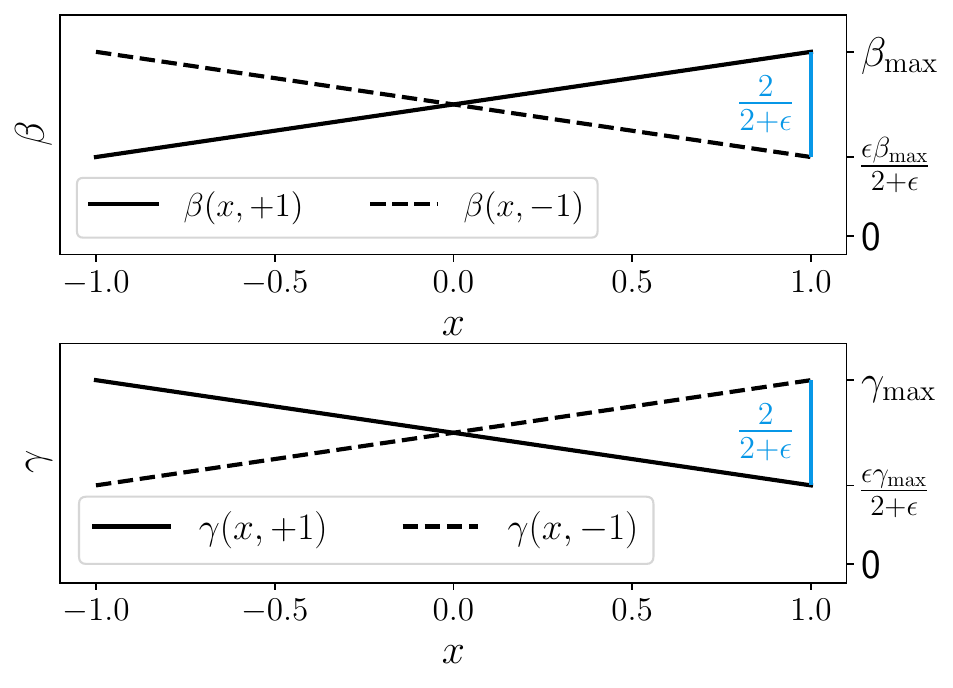}
    \caption{Infection  and recovery rates $\beta(x,s)$ (top) and $\gamma(x,s)$ (bottom)  as functions of the opinion $x$ for  parameters $\beta_{\text{max}}=1$, $\gamma_{\text{max}}=1$, and $\epsilon=1.5$, for $s = +1$ (solid lines) and $s = -1$ (dashed lines).}
    \label{infectionratefig}
\end{figure}

We assume that time evolves in discrete steps, $t=0,\,\Delta t,\, 2\Delta t,\, \dots$\,. A single time step of the agent-based model is as follows:
\begin{enumerate}
    \item $M$ nodes are selected uniformly at random to act as ``spreaders".
    \item For each spreader node $i$:
    \begin{enumerate}
        \item if $s^t_i=0$, nothing is done. Otherwise, one network neighbor $j$ of $i$ is selected uniformly at random to be exposed.
        \item The opinion of node $j$ is updated to
        \begin{eqnarray}
            x_j^{\, t+\Delta t}&=&x_{j}^{\, t} + C(s^t_i - x_{j}^{\, t})\Delta t\, ,\label{distime_dx}
        \end{eqnarray}
        where $C\in \mathbb{R}^{+}$ is the rate of opinion shift. If $x_j^{t+\Delta t}$ is larger than  $+1$ (less than $-1$), $x_j^{t+\Delta t}$ is set to $+1$ ($-1$).
        \item If $s^t_{j}=0$, node $j$ is infected with contagion $s_i^t$ (i.e., $s_j^{t+\Delta t}=s_i^t$) with probability
        \begin{eqnarray}
            \beta(x^t_{j}, s^t_i)\Delta t\, .
        \end{eqnarray}
    \end{enumerate}
    \item Each infected node, $d$, heals with probability
    \begin{eqnarray}
        \gamma(x^t_{d},s^t_d)\Delta t\, .
    \end{eqnarray}
\end{enumerate}
The infection and recovery rates are given, respectively, by
\begin{figure}[b]
    \centering
    \includegraphics[width=\linewidth]{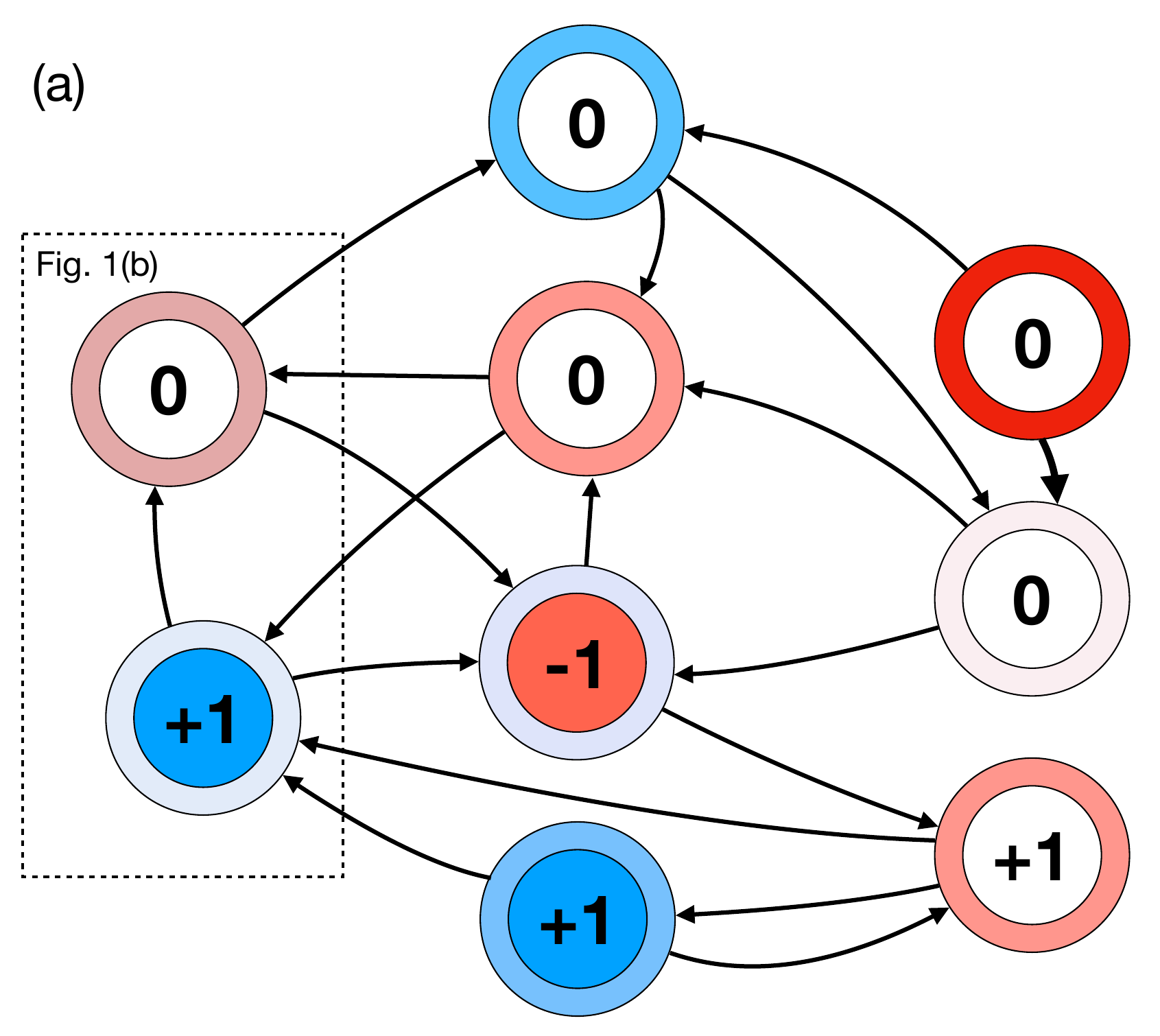}
    \includegraphics[width=\linewidth]{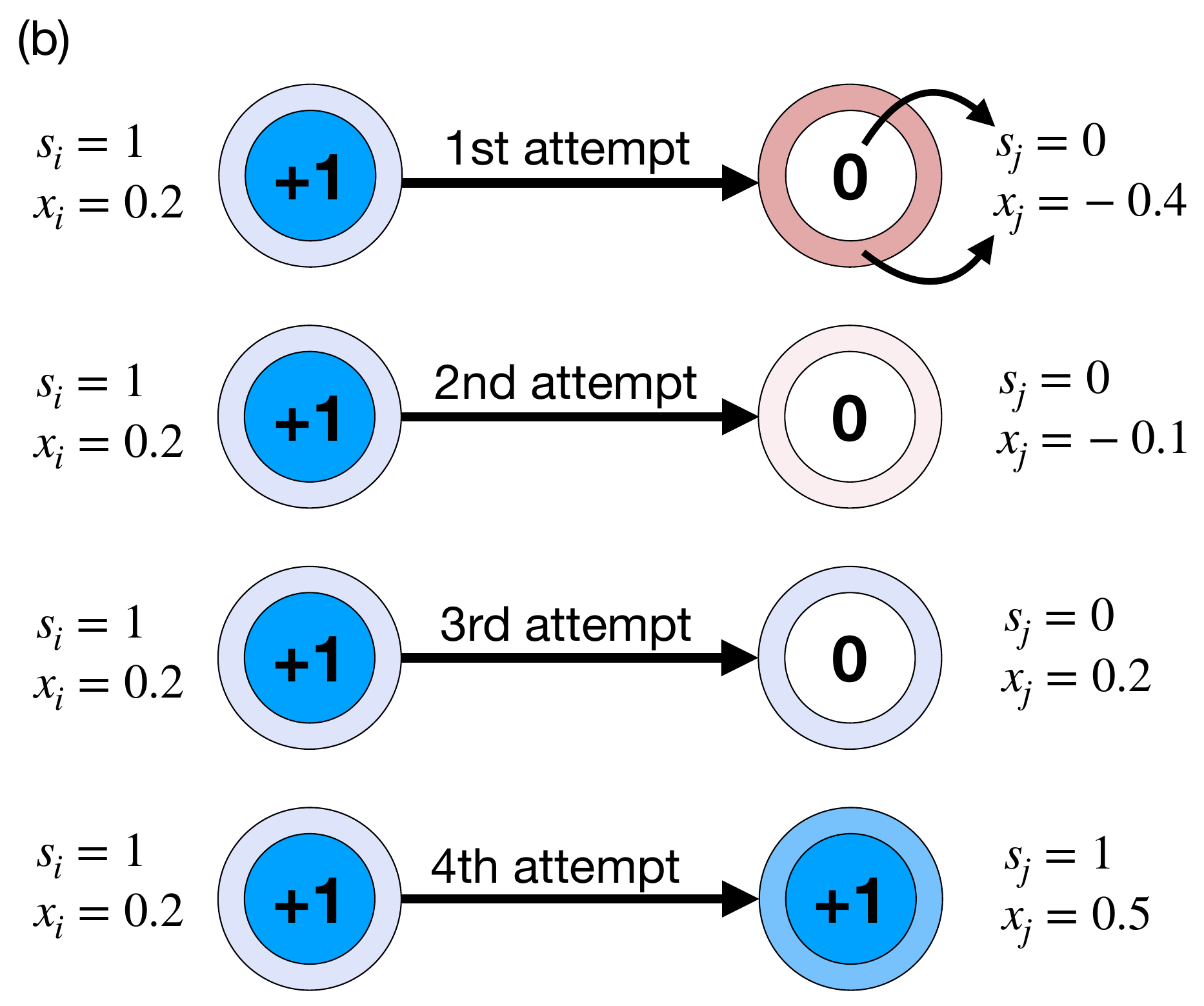}
    \caption{\label{diagram} (a) A diagram showing an example interaction network for the agent-based model. Each node has an contagion state $-1$ (red), 0 (white), or $+1$ (blue) (shown as the inner circle). The internal opinion of each node is represented by the color of the outer ring, ranging from $-1$ (red) to $+1$ (blue). (b) An example of how repeated exposures can change the opinion of node $j$ to align with the contagion of node $i$ and the possible transition of node $j$ from susceptible to infected with the $+1$ contagion.}
    \label{cartoonfig}
\end{figure}
\begin{eqnarray}
    \beta(x,s)&=& \frac{1+sx+\epsilon}{2+\epsilon}\,\beta_{\text{max}}\, ,\label{betabar}\\
    \gamma(x,s)&=&\frac{1-s x + \epsilon}{2+\epsilon}\,\gamma_{\text{max}}\, ,\label{gamma}
\end{eqnarray}
where $\epsilon>0$  is a parameter which measures the difference between the smallest and largest values of $\beta$ and $\gamma$ as shown in Fig.~\ref{infectionratefig}. The particular choice of $\beta$ in Eq.~(\ref{betabar}) was made so that the infection rate of a node $i$ with opinion $x_i$ to contagion $s$ is larger if $x_i$ is close to $s$, to model confirmation bias. Similarly, the form for $\gamma$ in Eq.~(\ref{gamma}) was selected such that $\gamma$ is smaller if $x_i$ is close to $s$, to model the unwillingness to give up an idea that the individual has a strong belief in. In addition, $\gamma$ increases as $x_i$ gets closer to $-s$, allowing individuals to stop spreading a contagion that is inconsistent with their views. 

The parameter $\epsilon$ controls the strength of the confirmation bias: for $\epsilon=0$, the infection and healing rates dependence on the node's internal opinion is the strongest; as $\epsilon\rightarrow \infty$, the infection and healing rates become independent of the node's internal opinion. 

In simulations of our agent-based model each node is assigned an initial internal opinion and an initial discrete contagion state. To set the initial conditions for the fractions of nodes in the three contagion states, subsets of the $n$ nodes are selected to be infected with the $+1$ contagion, $-1$ contagion, or left susceptible. These subsets are constructed by drawing $N_+$ agents uniformly at random from all agents to be infected with the $+1$ contagion. From the remaining agents, an additional $N_-$ are selected uniformly at random to be infected with the $-1$ contagion. The remaining $n-N_+-N_-$ agents are left susceptible. The initial internal opinions are assigned homogeneously (i.e., all nodes begin with the same internal opinion) in Sections IV and V, while heterogeneous distributions of initial opinions are explored in Section VI.

Fig.~\ref{diagram} illustrates our model. Fig.~\ref{diagram}(a) shows a network where each node has a contagion state which is either $-1$ (red), 0 (white), or $+1$ (blue), shown in the inner circle of each node. The opinion of each node is a continuous variable $x$ represented with the color of the outer circle of each node. Fig.~\ref{diagram}(b) shows an example of the illusory truth effect and confirmation bias in our model, as a node with contagion $s=+1$ and opinion $x=0.2$ (left) repeatedly attempts to infect another node (right). With each attempt, the opinion of the node on the right gets closer to $+1$ (illusory truth effect), thus making the node more susceptible to the contagion (confirmation bias).

\section{\label{MFA}Mean-field Approximation}

To study the dynamics of this model we develop a mean-field approximation for the dynamics of the average opinion and the fraction of nodes with the $+1$ and $-1$ contagions, given respectively by 
\begin{eqnarray}
    X&=&\frac{1}{N}\sum_{i=1}^Nx_i\, ,\label{OX}\\
    S_+&=&\frac{1}{N}\sum_{s_i=1}s_i\, ,\label{OSP}\\
    S_-&=&\frac{1}{N}\sum_{s_i=-1}|s_i|\, .\label{OSN}
\end{eqnarray}
For simplicity, we develop our mean-field approximation only for $k$-regular networks. The expected fraction of nodes that recover from $\pm 1$ contagion in a small time step of length $\Delta t$ is approximately 
\begin{eqnarray}
    \gamma(X, \pm 1)S_{\pm}\,\Delta t\, .
\end{eqnarray}
Similarly, the expected fraction of nodes that become infected with the $\pm 1$ contagion in a small time step of length $\Delta t$ is approximately 
\begin{eqnarray}
    \frac{M}{N}S_{\pm}\frac{k-1}{k}(1-S_{+} - S_{-})\beta(X, \pm 1)\,\Delta t\, ,
\end{eqnarray}
where $MS_{\pm}$ is the expected number of spreader nodes with the $\pm 1$ contagion, $\frac{k-1}{k}(1-S_+-S_-)$ is the probability that the randomly chosen neighbor of the spreader node is susceptible, and $\beta(X,\pm 1)\Delta t$ is the probability that the spreader node successfully infects the susceptible neighbor. To understand the need for the factor $(k-1)/k$, note that $(1-S_+-S_-)$ would be the expected fraction of susceptible neighbors of the spreader node if these nodes were selected uniformly at random. However, this estimate neglects the fact that neighbors of a spreader node are not chosen uniformly at random, but their choice is conditioned on being neighbors of an already infected node. Since the spreader node must have been infected by one of its neighbor nodes, we remove one node from the count by multiplying by the factor $(k-1)/k$ (this first-order correction neglects the possibility that the node might have healed since it infected the spreader node). 

From Eq.~(\ref{distime_dx}) the average change in opinion over the small time interval $\Delta t$ due to attempted infections from nodes with the $\pm 1$ contagion is approximately
\begin{eqnarray}
    \frac{M}{N}C(\pm 1 - X)S_{\pm}\,\Delta t \,.
\end{eqnarray}\\
In the limit $\Delta t \rightarrow 0$ these approximations result in the following system of differential equations for the three order parameters in Eqs.~(\ref{OX})-(\ref{OSN}):
\begin{eqnarray}
    \frac{dS_{+}}{dt}&=&-S_{+}\gamma(X, 1) \notag \\
    &+& \frac{M}{N} \frac{k-1}{k}(1 - S_{+} - S_{-})S_{+}\beta(X, 1), \label{SPfull} \\
    \frac{dS_{-}}{dt}&=&-S_{-}\gamma(X, -1)\notag \\
    &+& \frac{M}{N} \frac{k-1}{k}(1 - S_{+} - S_{-})S_{-}\beta(X, -1), \label{SNfull} \\
    \frac{dX}{dt}&=& \frac{M}{N}C\left[(1-X)S_{+} - (1+X)S_{-} \right], \label{OPfull}
\end{eqnarray}
where $\beta$ and $\gamma$ are given by Eqs. (\ref{betabar}) and (\ref{gamma}). Substituting Eqs.~(\ref{betabar}) and (\ref{gamma}) and non-dimensionalizing Eqs.~(\ref{SPfull})-(\ref{OPfull}) we arrive at the reduced equations
\begin{eqnarray}
    \frac{dS_+}{d\tau}&=&-S_+\left(1-X+\epsilon \right) \notag \\ &+&r_0(1-S_+-S_-)S_+\left(1+X+\epsilon\right)\, , \label{SP} \\
    \frac{dS_-}{d\tau}&=&-S_-\left(1+X+\epsilon\right) \notag \\ &+&r_0(1-S_+-S_-)S_-\left(1-X+\epsilon\right)\, , \label{SN}\\
    \frac{dX}{d\tau}&=&K\left[(S_+-S_-)-(S_+ +S_-)X \right]\, , \label{OP}
\end{eqnarray}
where $r_0$, $\tau$, and $K$ are defined as 
\begin{eqnarray}
    r_0&=&\frac{M\beta_{\text{max}}(k-1)}{Nk\gamma_{\text{max}}}\, ,\label{r0}\\
    \tau&=&\frac{\gamma_{\text{max}}}{2+\epsilon}\, ,\\
    K&=& MC/N\, .
\end{eqnarray}

Note that Eqs.~(\ref{SP}) and (\ref{SN}) correspond to the SIS model for a pair of competing contagions where the healing and infection rates are $[1-X+\epsilon]$ and $r_0[1+X+\epsilon]$, respectively, for the $+1$ contagion, and $[1+X+\epsilon]$ and $r_0[1-X+\epsilon]$ for the $-1$ contagion. The healing and infection rates are controlled by the average opinion $X$, which in turn depends dynamically on the fraction of infected individuals, $S_+$ and $S_-$, via Eq.~(\ref{OP}). In the next section we will study the conditions under which one contagion becomes prevalent while the other disappears. First, however, we discuss some of the assumptions made in developing the model and its mean-field description.

Our mean-field description is based on the assumption of a homogeneous network where each node has degree $k$. However, our analysis could be extended to networks with heterogeneous degree distributions using the methods of Ref.~\cite{istvan2017mathematics}. We have conducted our numerical simulations of the agent-based model using target $k$-regular networks constructed via the configuration model and found good agreement with our mean-field approximation even for $k$ as small as $7$ (see Fig.~\ref{TS_rebound}). The target $k$-regular networks used in this project were constructed using the complex group interactions (XGI) package for Python \cite{Landry_XGI_2023}. In addition, our mean field description neglects pair correlations \cite{istvan2017mathematics}.

Other assumptions of our model are the particular functional forms for how the healing and infection rates depend on a node's opinion, and how the opinion changes upon an attempted infection. We chose the forms in Eqs.~(\ref{distime_dx}), (\ref{betabar}), and (\ref{gamma}) for simplicity, and we expect qualitatively similar results for other choices where the infection rates are increasing and decreasing function of the node's opinion for the $+1$ and $-1$ contagions, respectively, and vice versa for the healing rates. 

Our model is based on sequential (rather than simultaneous) updating. The basic update rule in our model is the selection of a random node which, if infected, attempts to spread the contagion. In order to speed up the numerical simulation of our model, $M$ such updates are carried out every time step. Alternatively, one could consider a simultaneous updating version of our model, where on every time step every infected node attempts to spread its contagion with a certain probability. Although we have not explored this version of our model, it exists as a special case where $M=n$ (i.e., selecting every node at each time step). 

When comparing the agent-based model and the mean-field approximation, it is necessary to match their initial conditions. To achieve this, given initial conditions $S_{\pm}(0)$ for the initial fraction of spreaders in the mean-field model, we compute the initial number of spreaders for each contagion in the agent-based model as $N_{\pm}=\text{floor}[N S_{\pm}(0)]$. For the opinions, given a mean opinion $X(0)$ in the mean-field model, we sample their initial opinion from a distribution with mean $X(0)$. In Secs.~IV and V we assign all agents in the agent-based model the same opinion $X(0)$. In Sec.~VI we allow agents to start with different opinions by sampling their initial opinion from a Gaussian and a bimodal distribution. With these choices, the average opinion of the agents in the agent-based model will be $X(0)$ and the initial fraction of spreaders of the $+1$ and $-1$ contagions will be approximately $S_+(0)$ and $S_-(0)$, respectively. This allows us to specify the initial conditions of both the agent-based model and mean-field equations as the ordered triplet $(X(0),S_+(0),S_-(0))$. Note that these choices do not allow the initial conditions for opinions and contagion states to be correlated, a potentially important case that will be studied in future work.

\section{\label{STB} Equilibria and their Stability}

The mean-field equations (\ref{SP})-(\ref{OP}) admit the following equilibrium solutions:
\begin{itemize}
    \item {\it Disease-free behavior}: The family $(X, S_{+}, S_{-})=(A,0,0)$, where $A$ is an arbitrary constant. This family corresponds to the case where both contagions are absent, but there is an underlying average opinion $X=A$.
    \item {\it Endemic behavior}: The two equilibria $(X, S_{+}, S_{-})=(+1,S^*,0)$ and $(X, S_{+}, S_{-})=(-1,0,S^*)$, where 
    \begin{eqnarray}
        S^*&=&1-\frac{\epsilon}{(2+\epsilon)r_0}\, .
    \end{eqnarray}
    These two equilibria correspond to the case where one contagion drives the other one to extinction, and the surviving contagion drives the average opinion to consensus (here understood as the state where all nodes hold the same opinion). We refer to these cases, respectively, as {\it +1 endemic behavior} and {\it -1 endemic behavior}.
    \item {\it Coexistence}: The equilibrium point $(X,S_+,S_-)=(0,\sigma,\sigma)$, where 
    \begin{eqnarray}
        \sigma&=&\frac{r_0-1}{2r_0}\, .
    \end{eqnarray}
    This equilibrium point corresponds to a case where the two contagions coexist and the average opinion is zero. However, linear stability analysis shows that this solution is unstable.
\end{itemize}

 The local stability of the disease-free equilibrium solutions $(A,0,0)$ depends on the condition that $R_{\, 0}^{\, \pm}(A)<1$, where the effective reproduction numbers $R_0^+(A)$ and $R_0^{-}(A)$ are given by
\begin{eqnarray}
    R_{\,0}^{\,+}(A)&=&r_0\left[\frac{1+A+\epsilon}{1-A+\epsilon} \right]\, , \label{R0p1}\\
    R_{\, 0}^{\, -}(A)&=&r_0\left[\frac{1-A+\epsilon}{1+A+\epsilon} \right]\, \label{R0n1}.
\end{eqnarray}
 Similarly, a linear stability analysis about $(+1,S^*,0)$ and $(-1,0,S^*)$ results in the conditions $R^{+}_0(A)> 1$ and $R^{-}_0(A)> 1$, respectively, for these points to be stable. 

 For a given $r_0$ and $\epsilon$, the value of the average opinion $A$ that results in instability of the disease-free state towards the $+1$ or $-1$ contagions (i.e., such that the unstable manifold of the disease-free state is a subset of the basin of attraction of the $+1$ or $-1$ endemic state) can be found by setting $R^{+}_0$ or $R^{-}_0$ equal to $1$. When $R^{+}_0=1$, Eq.~(\ref{R0p1}) gives
\begin{eqnarray}
    A^+&=&\text{max}\left\{ (1+\epsilon)\left[\frac{1-r_0}{1+r_0} \right], 0 \right \}\,. \label{AP}
\end{eqnarray}

Similarly, setting $R_0^{-}=1$ we get
\begin{eqnarray}
    A^-&=&\text{min}\left\{ -(1+\epsilon)\left[\frac{1-r_0}{1+r_0} \right], 0 \right \}\,, \label{AM}
\end{eqnarray}
from Eq.~(\ref{R0n1}). The inclusion of the $\text{max}(\cdot)$ and $\text{min}(\cdot)$ functions in Eqs.~(\ref{AP}) and (\ref{AM}), respectively, is to ensure that $A^+\geq 0$ and $A^-\leq 0$ for values of $r_0>1$. This is done to emphasize that $+1$ endemic behavior and $-1$ endemic behavior cannot simultaneously be stable.

The values (\ref{AP}) and (\ref{AM}) provide bounds on the average opinion for which each of the three equilibria are stable. Particularly, for $A\in (A^-,A^+)$ the equilibrium $(A,0,0)$ is stable. For $A> A^+$, $(+1,S^*,0)$ is stable, and for $A< A^{-}$ the equilibrium $(-1,0,S^*)$ is stable. Notice that for any value of $A$, there is always exactly one stable equilibrium point.

 \begin{figure}[t]
    \includegraphics[width=\linewidth]{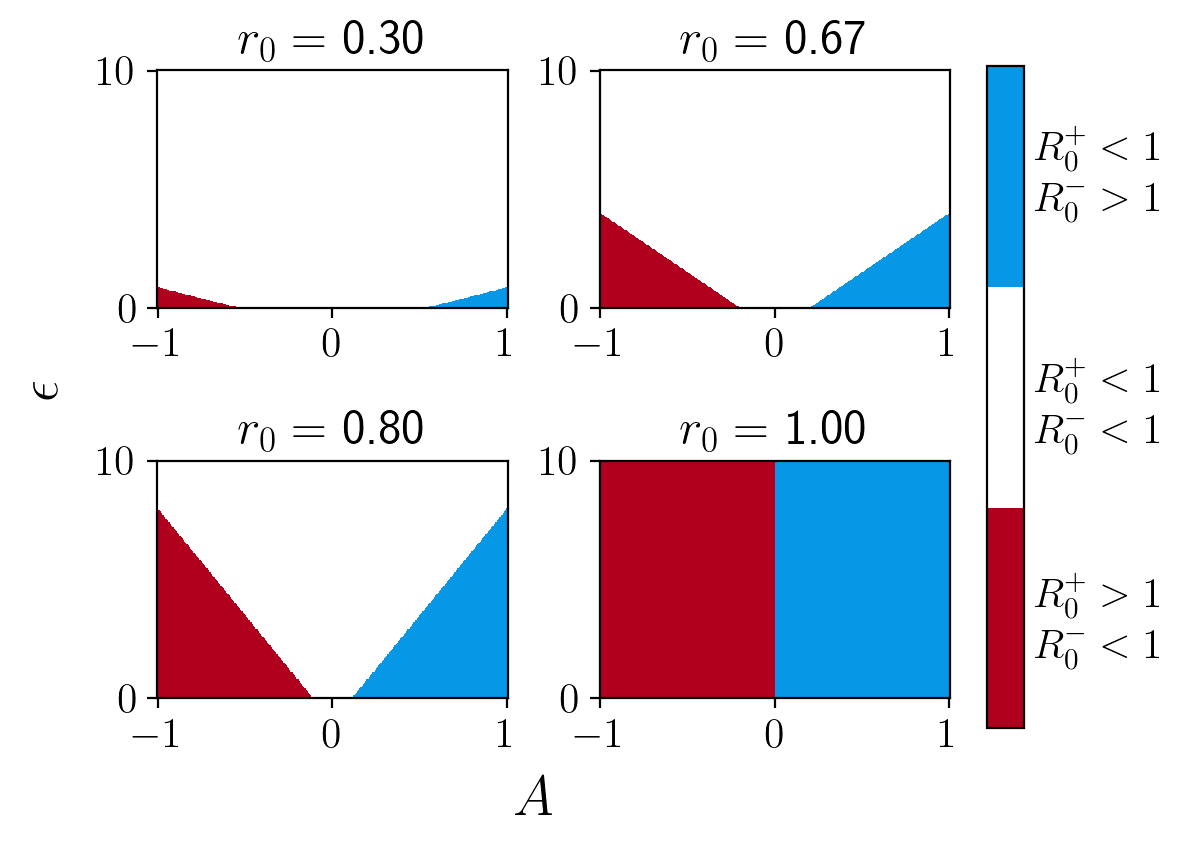}
    \caption{\label{stability} Examples of the stability of the $(A,0,0)$ equilibrium state, from Eqs.~(\ref{R0p1}) and (\ref{R0n1}), for values of $r_0\in \{0.30,0.67,0.80,1.00 \}$ as a function of $A$ and $\epsilon$. Red represents unstable towards the $-1$ contagion ($R_0^+<1$ and $R_0^->1$), blue represents unstable towards the $+1$ contagion ($R_0^+>1$ and $R_0^-<1$), and white represents stable ($R_0^+<1$ and $R_0^-<1$).}
\end{figure}
In Fig.~\ref{stability} we show how the stability of the disease-free equilibrium $(A,0,0)$ depends on $A$ and $\epsilon$ for values of the reproductive number $r_0$ given by $0.3$ (top left), $0.67$ (top right), $0.8$ (bottom left), and $1$ (bottom right). In each panel, the color white indicates stability of the disease-free state [i.e., $R_0^+(A)<1$, $R_0^-(A)<1$], red indicates instability towards the $-1$ contagion [$R_0^-(A)>1$, $R_0^+(A)<1$], and blue instability towards the $+1$ contagion [$R_0^+(A)>1$, $R_0^-(A)<1$]. 

\begin{figure}[t]
    \includegraphics[width=\linewidth]{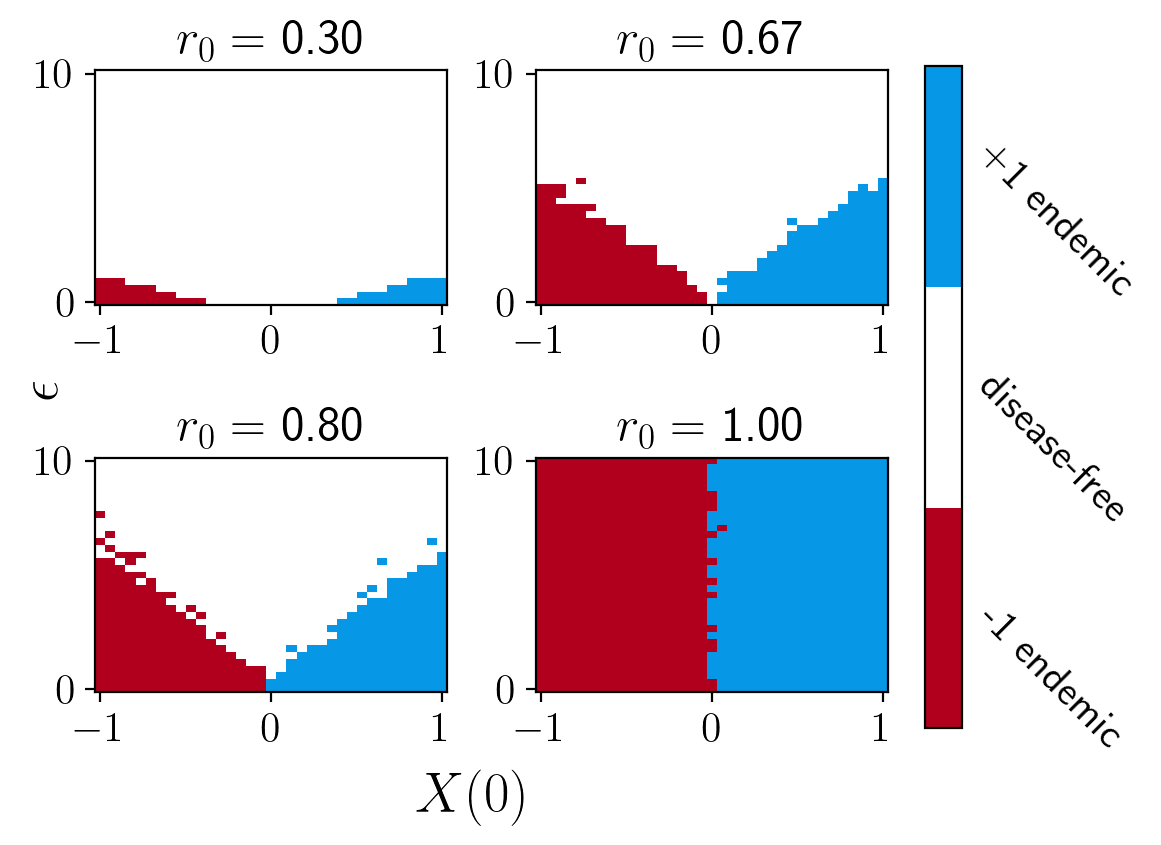}
    \caption{\label{simstability} The long-term behavior of the agent-based model for $r_0\in \{0.30,0.67,0.80,1.00 \}$ as a function of $X(0)$ and $\epsilon$. The color of each point represents which contagion was successful more frequently out of $9$ independent trails of the agent-based simulation on a $30$-regular network of $N=1000$ nodes. Each simulation ran for 3000 time steps with initial fractions of infected nodes given as $(S_+(0),S_-(0))\in \{(0.05i/2,\, 0.05j/2)|\,i\in \{0,1,2\},\,j\in \{0,1,2\} \}$.}
\end{figure}

Fig.~\ref{simstability} shows the results obtained from numerical simulation of the agent-based model for a target $k$-regular network with $N=1000$ and $k=30$ and the same values of $r_0$ shown in Fig.~\ref{stability}. For each choice of $(X(0),\epsilon)$, the agent-based model was simulated $9$ times with initial fractions of infected nodes $(S_+(0),S_-(0))$ spaced uniformly in the square $[0,0.05]\times [0,0.05]$ [i.e., $(S_+(0),S_-(0))\in \{(0.05i/2,\, 0.05j/2)|\,i\in \{0,1,2\},\,j\in \{0,1,2\} \}$]. After 3000 time steps the fraction of nodes with each contagion was stored. After the $9$ independent simulations for each pair $(X(0),\epsilon)$ the mean final fraction of nodes with each contagion across the $9$ simulations was computed. When the mean final fraction of nodes with the $+1$ contagion was larger than the mean final fraction of nodes with the $-1$ contagion, a $1$ was recorded (blue). Conversely, when the mean final fraction of nodes with the $-1$ contagion was larger than the mean final fraction of nodes with the $+1$, a $-1$ was recorded (red). Otherwise, a zero was recorded (white). The choice of 3000 time steps allows sufficient time for the system to converge such that the average opinion at 3000 time steps deviates from the mean over the last 200 time steps by less than $0.001$. We will use the same number of time steps in the rest of the paper unless noted. The same condition is fulfilled in the rest of the simulations. Overall, the mean-field approximation and the numerical simulations of the agent-based model agree well for $k$-regular networks for $k = 30$. We also observe reasonably good agreement for $k$ as small as $7$ (cf. Fig.~\ref{TS_rebound}).

\section{\label{behavior}Rebound and Bias Overturning}

In the previous section we found that the mean-field version of our model admits disease-free and endemic states, where stability is dependent on the average opinion. Since the average opinion is a dynamic quantity, the transient and long-term behavior of our model depend in a non-trivial way on the initial conditions. Two examples of the complex dependence of the final state on the initial conditions are the {\it rebound} and the {\it bias overturning} behaviors, which we discuss below.

\begin{figure}[t]
\includegraphics[width=\linewidth]{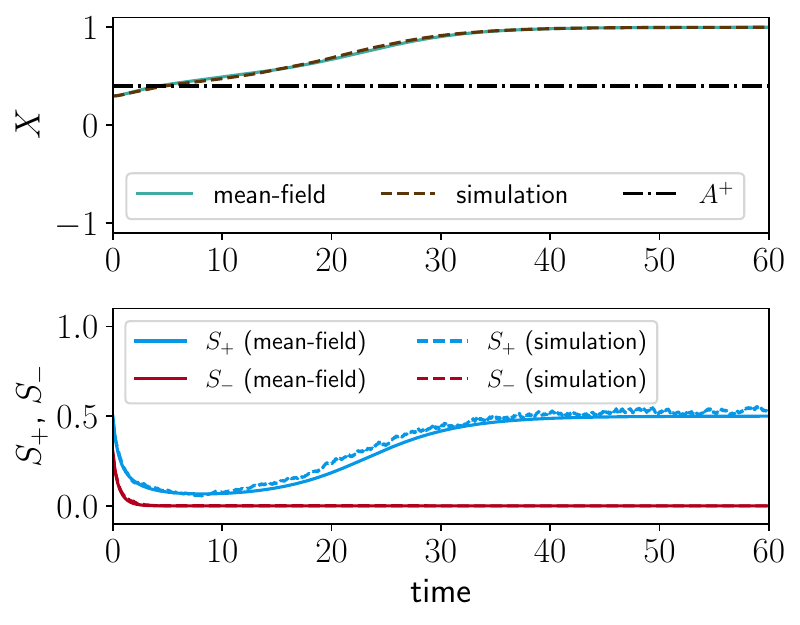}
    \caption{\label{TS_rebound} An example of a rebound to the $+1$ endemic state for a single simulation of the agent-based model and single numerical solution to the mean-field equations~(\ref{SP})-(\ref{OP}) with parameters $r_0=0.66$, $K=0.4$, and $\tau=0.07$ on a 7-regular network. The dot-dashed black line indicates the threshold $A^+$ from Eq.~(\ref{AP}).}
\end{figure}

\begin{figure}[t]
\includegraphics[width = \linewidth]{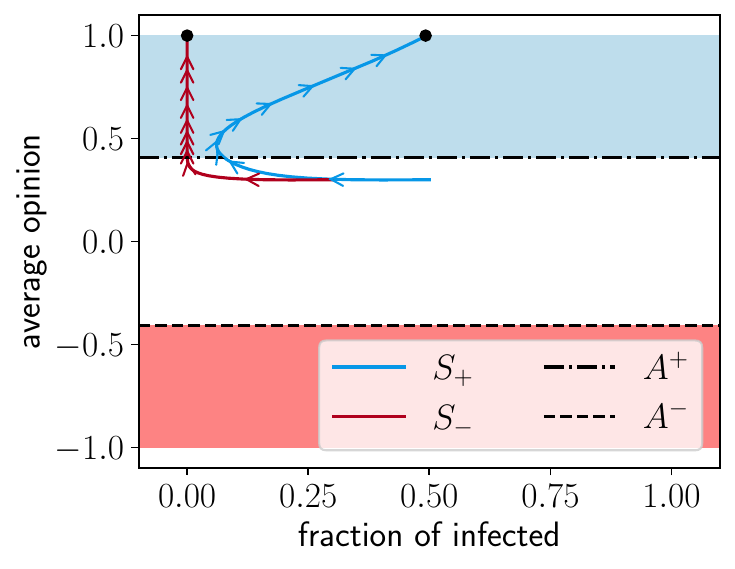}
    \caption{\label{PS_rebound} A phase space diagram showing an example of a rebound to the $+1$ endemic state from a single numerical solution to the mean-field equations~(\ref{SP})-(\ref{OP}) for $r_0=0.66$, $K=0.4$, and $\tau=0.07$ on a 7-regular network. The blue and red lines with arrows indicate $(S_+,X)$ and $(S_-,X)$ trajectories, respectively. The black dots indicate the equilibria of the $+1$ and $-1$ contagions and the dot-dashed and dashed black lines indicate the thresholds $A^+$ and $A^-$ from Eqs.~(\ref{AP}) and (\ref{AM}) respectively.}
\end{figure}

In the rebound, the initial conditions $(X(0),S_+(0),S_-(0))$ are such that $A^-<X<A^+$, so that only the disease-free state with $S_+=0=S_-$ would be stable if $X$ was constant. As $S_+$ and $S_-$ decay to zero, $X$ changes and moves out of the interval $[A^-,A^+]$, thus bringing the system into the basin of attraction of either the $+1$ endemic or $-1$ endemic states, depending on whether $X>A^+$ or $X<A^-$, respectively. Fig.~\ref{TS_rebound} shows an example of a rebound. Fig.~\ref{TS_rebound}(a) shows the average opinion $X$ obtained from the agent-based model (teal solid line) and from the mean-field model (brown dashed line). Fig.~\ref{TS_rebound}(b) shows $S_+$ and $S_-$ obtained from the agent-based model (solid lines) and from the mean-field model (dashed lines). As discussed above, while both $S_+$ and $S_-$ initially decay, $X$ increases, at some point exceeding $A^+$ (black dot-dashed line). Subsequently, $S_+$ increases while $S_-$ keeps decaying. This is further illustrated in Fig.~\ref{PS_rebound}, which shows the trajectories of $(S_+,X)$ and $(S_-,X)$. After the trajectories enter the region where $X>A^+$ (blue region), they converge to the $+1$ endemic state equilibrium $(S^*,1)$ and $(0,1)$ (black circles). 

\begin{figure}[t]
\includegraphics[width = \linewidth]{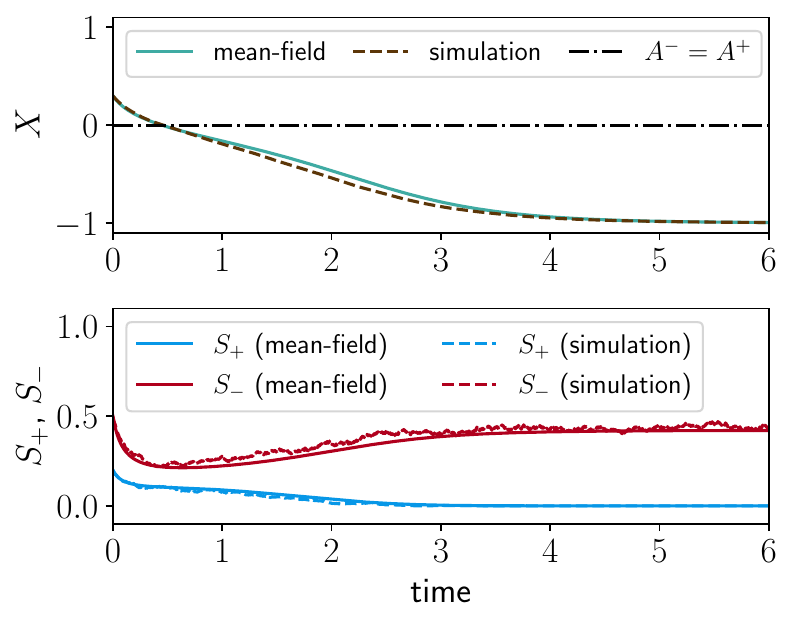}
    \caption{\label{TS_overturn} An example bias overturning for a single simulation of the agent-based model and a single numerical solution to the mean-field equations Eqs.~(\ref{SP})-(\ref{OP}) with parameters $r_0=1.44$, $K=3.20$, and $\tau=0.017$ on a 10-regular network. The dot-dashed black line indicates the threshold $A^+=A^-$ from Eqs.~(\ref{AP}) and (\ref{AM}).}
\end{figure}

\begin{figure}[t]
\includegraphics[width = \linewidth]{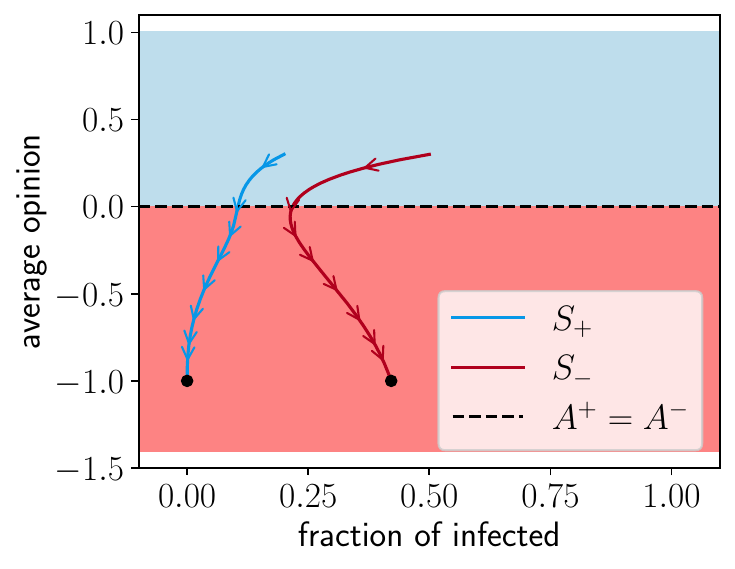}
    \caption{\label{PS_overturn} A phase space diagram showing an example bias overturning in the mean-field equations Eqs.~(\ref{SP})-(\ref{OP}) with parameters $r_0=1.44$, $K=3.20$, and $\tau=0.017$ on a 10-regular network. The blue and red lines with arrows indicate $(S_+,X)$ and $(S_-,X)$ trajectories, respectively. The black dots indicate the equilibria of the $+1$ and $-1$ contagions and the dashed black line indicates the threshold $A^+=A^-$ from Eqs.~(\ref{AP}) and (\ref{AM}).}
\end{figure}

The bias overturning behavior is characterized by the sign of the average opinion in the final state being opposite of that in the initial state. An example is shown in Figs.~\ref{TS_overturn} and \ref{PS_overturn}, with the same conventions as those used in Figs.~\ref{TS_rebound} and \ref{PS_rebound}. As shown in Fig.~\ref{TS_overturn}, the initial value of $X$ is positive. However, there is an excess of spreaders for the $-1$ contagion which, even as their numbers decay, manage to make $X$ negative, crossing $A^-$ (dot-dashed line), causing the system to converge to the $-1$ endemic state.  

\begin{figure}
\includegraphics[width = \linewidth]{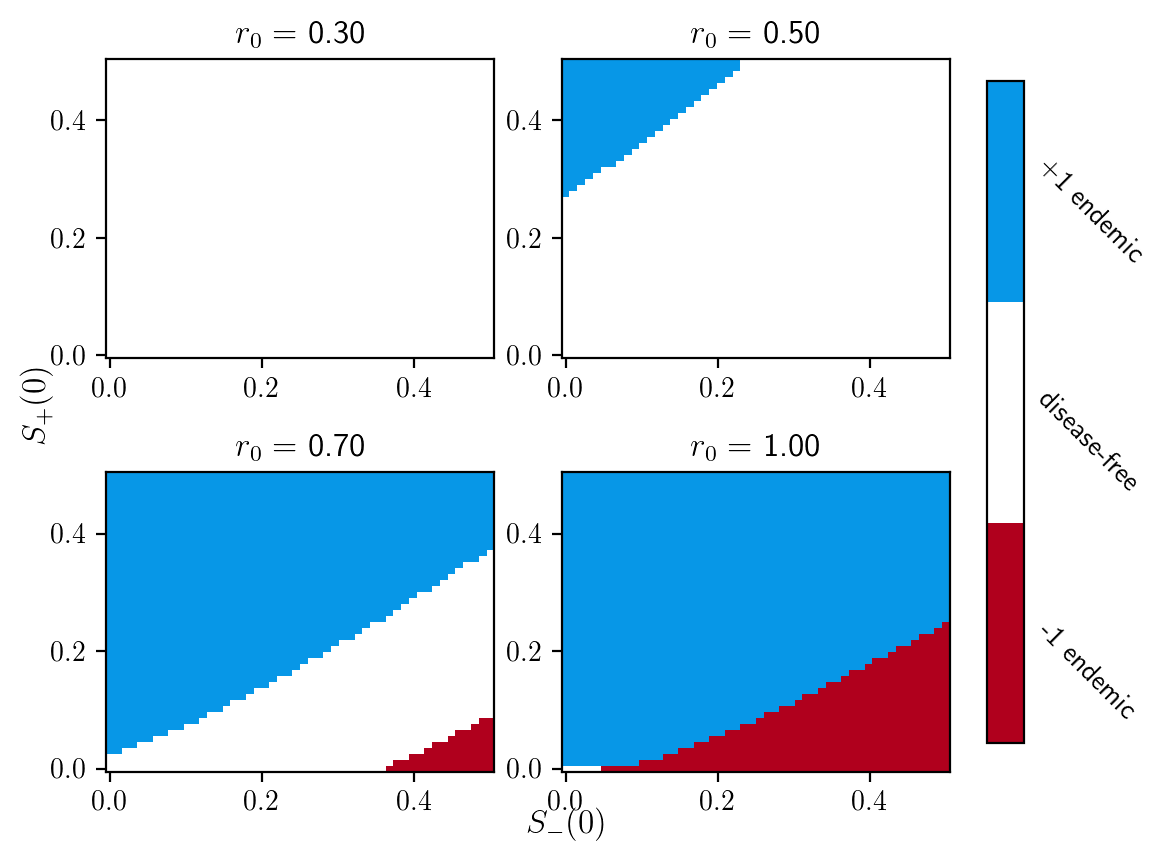}
    \caption{\label{IC_heatmap} Long-term behavior of the mean-field equations~(\ref{SP})-(\ref{OP}) after 3000 time steps as a function of the initial fraction of infected individuals in states $S_+(0)$ and $S_-(0)$ for $r_0\in \{0.30,0.50,0.70,1.0\}$, $K=0.25$, $\tau=0.1$, $X(0)=0.1$, and a time step of size $h=0.25$.}
\end{figure}

These examples illustrate how the final state of the system depends on the initial values of $S_+$, $S_-$, and $X$. To illustrate this in a more systematic way, in Fig.~\ref{IC_heatmap} we have plotted the basin of attraction of the $+1$ endemic (blue), $-1$ endemic (red), and disease-free (white) cases for $S_+(0),S_-(0)\in[0,0.5]$ and $r_0\in \{0.3,0.5,0.7,1.0 \}$ with an initial opinion bias $X(0)=0.1$, obtained using  numerical solutions of the mean-field equations (\ref{SP})-(\ref{OP}) simulated over $t_f=3000$ time steps. There we see that as $r_0$ increases the disease-free region becomes smaller until the system transitions directly between the $+1$ and $-1$ endemic cases (lower right panel). Again, since $X$ always changes towards the dominant contagion, if the system is near the transition boundaries (\ref{AP}) and (\ref{AM}) then a sufficiently large initial portion of the population infected with the opposite contagion can result in the initial bias of the population being overturned. Fig.~\ref{IC_heatmap} corresponds to an initial mean opinion $X(0)=0.1$. When a different value of $X(0)$ is used, the basins of attraction for each steady-state grow/shrink in accordance. For example, using a more positive value makes the basins of attraction for the $+1$ endemic state larger, while the basins of attraction for the $-1$ endemic state and disease-free state become smaller. Additionally, we note that, for the parameters that we have studied, the bias overturning behavior only occurs when the initial bias is small.

When the disease-free region is quite large and bias overturning is impossible (e.g., Fig.~\ref{IC_heatmap} top right) having a sufficiently large initial population in the state opposite the initial average opinion can still push the system from either the $+1$ or $-1$ endemic states into the disease-free state. This behavior may have potential consequences for the spread of disinformation, as it suggests that artificially boosting the initial number of spreaders through, for example, social-bot networks may be sufficient to overcome an initial bias towards a belief and potentially sway public opinion. However, we have also observed that the average initial opinion has a strong effect on the possibility of bias overturning. This suggests that a suitably large initial bias, perhaps through prebunking efforts (the practice of proactively exposing and debunking misinformation before it spreads, see \cite{lewandowsky2021countering}) may provide good protection against bias overturning in the unforced system. We will be considering the effects of external forcing in Sec.~\ref{reintroduction}. It should also be noted that this  has implications both for protection against misinformation as well as the creation of misinformation, since it shows that if the system becomes initially biased towards a piece of false information, it may be quite difficult to overturn the bias in favor of the truth.

\section{\label{Opinion Distrobution} Heterogeneous Initial Opinions}

Until this point we have made the assumption that all agents begin with the same initial opinion. This assumption is highly unrealistic for any real-world system. In this section we investigate the effects of heterogeneous initial opinions on the behavior of our model and on the accuracy of our mean-field approximation. We consider two cases for the distribution of heterogeneous initial opinions. In the first case, we  consider opinions sampled from a normal distribution with mean $\mu$ and standard deviation $\sigma$ truncated to the interval $[-1,1]$, i.e., values which fall outside of this interval are redrawn from the same distribution until all values are within the given interval. In the second case, we consider an evenly weighted bimodal Dirac delta distribution with mean $\mu$ and a distance $\Delta$ between the two modes, i.e., a distribution given by $P(X) = [\delta(X - \mu +\Delta/2) + \delta(X- \mu -\Delta/2)]/2$. 

\begin{figure}
    \centering
    \includegraphics[width=\linewidth]{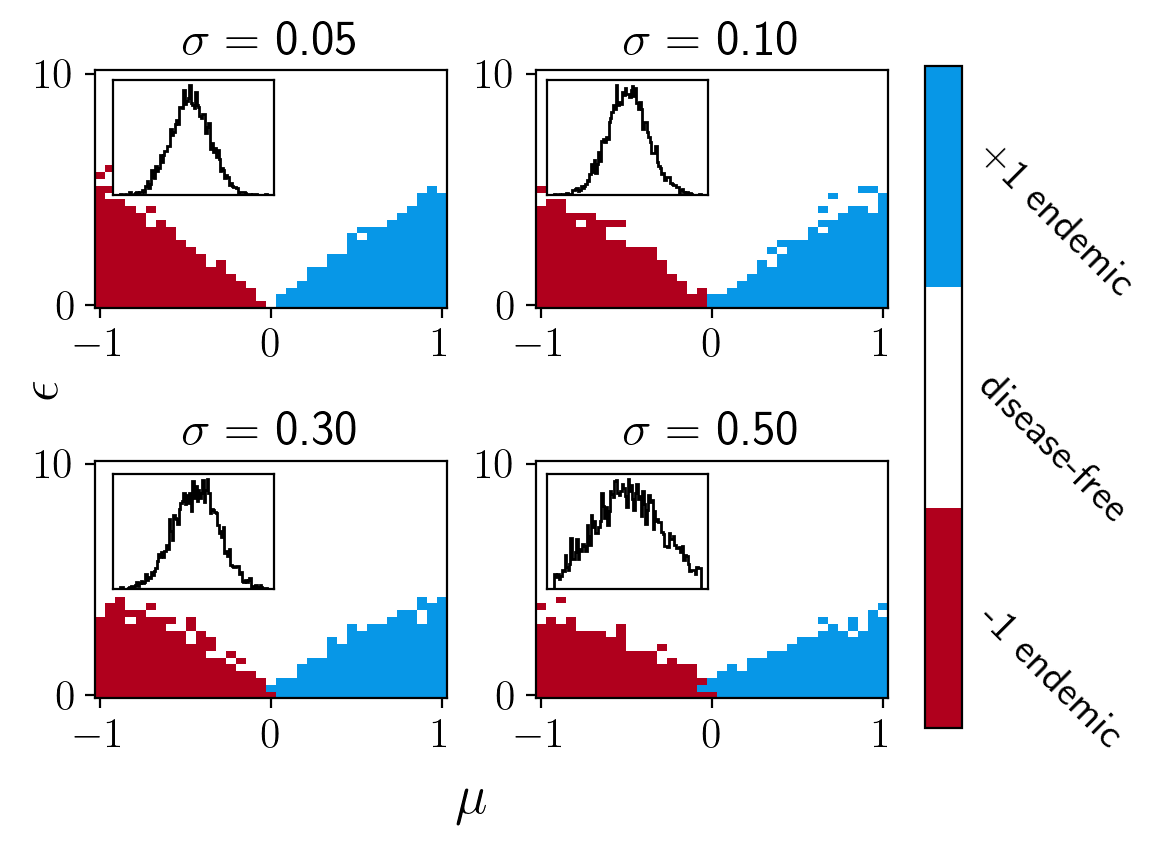}
    \caption{ The long-term behavior of the agent-based model for $r_0=0.67$ and $\sigma\in \{0.05,0.10,0.30,0.50 \}$ as a function of $\mu$ and $\epsilon$. The color of each point represents which contagion was successful more frequently out of 9 independent trails of the agent-based simulation on a 30-regular network of $N=1000$ nodes. Each simulation ran for $3000$ time steps with initial fraction of infected nodes given as $(S_+(0),S_-(0))\in \{(0.05i/2,\, 0.05j/2)|\,i\in \{0,1,2\},\,j\in \{0,1,2\} \}$. Insert shows a sample truncated Gaussian with mean $\mu$ for each value of $\sigma$.}
    \label{het_stability}
\end{figure}

Fig.~\ref{het_stability} shows the results obtained from numerical simulation of the agent-based model for a 30-regular network with $N=1000$ nodes with $r_0=0.67$ for the truncated normal distribution with $\sigma\in \{0.05,0.1,0.3,0.5\}$. For each pair $(\mu,\epsilon)$ the same procedure is used as in Fig.~\ref{simstability} with the same selection of $(S_+(0),S_-(0))$. The panels of Fig.~\ref{het_stability} should be compared to the  top-right panel of Fig.~\ref{simstability} ($r_0=0.67$). We observe that the general shape of the basins of attraction for the three possible steady state behaviors is qualitatively similar for smaller values of $\sigma$. As $\sigma$ increases, the similarity decreases; in particular, the basins of attraction of the endemic states become smaller. This shows that while our mean-field results are reasonably accurate with the inclusion of heterogeneous initial opinions, as the amount of heterogeneity is increased, this accuracy can be lost. Furthermore,  we notice that increasing $\sigma$ causes the basins of attraction of the $+1$ and $-1$ endemic cases to become smaller, which means that the disease-free state can be stable for smaller values of $\epsilon$. This indicates that when the initial distribution of opinions is more heterogeneous the disease-free state can remain stable even when the infection and healing rates are more strongly affected by an individuals opinion, suggesting that the role of confirmation bias and the illusory truth effect in the spread of information is smaller when the initial opinions are highly heterogeneous.

\begin{figure}
    \centering
    \includegraphics[width=\linewidth]{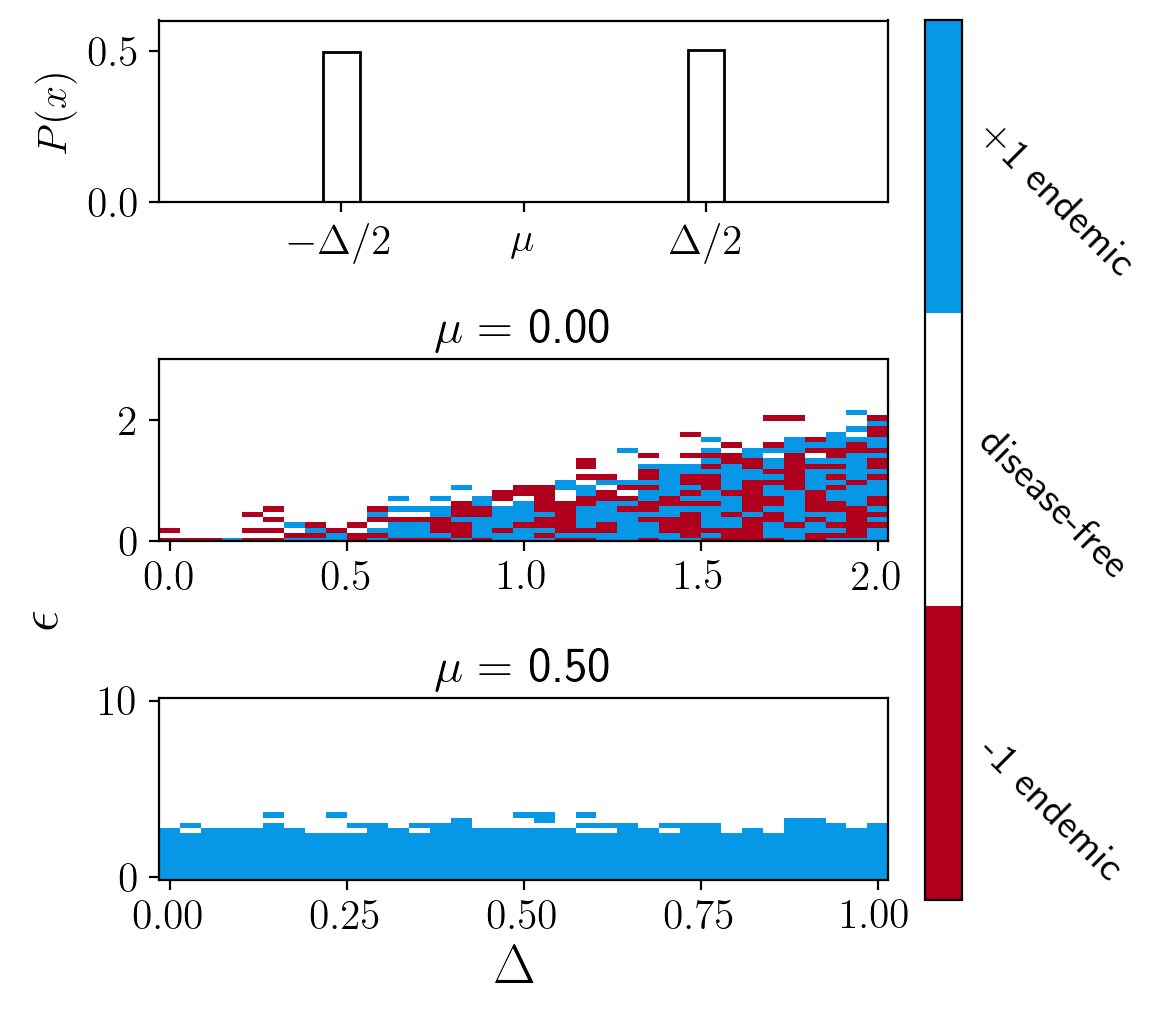}
    \caption{ (top) An example of the Dirac delta bimodal distribution with mean $\mu$ and distance $\Delta$ between modes. (middle) the long-term behavior of the agent-based model for $r_0=0.67$ with mean $\mu=0$ as a function of $\Delta$. (bottom) the long-term behavior of the agent-based model for $r_0=0.67$ with mean $\mu=0.5$ as a function of $\Delta$. The color of each point represents which contagion was successful more frequently out of the 9 independent trails of the agent-based simulation on a 30-regular network of $N=1000$ nodes. Each simulation ran for $3000$ time steps with initial fraction of infected nodes given as $(S_+(0),S_-(0))\in \{(0.05i/2,\, 0.05j/2)|\,i\in \{0,1,2\},\,j\in \{0,1,2\} \}$.}
    \label{bimodal_stability}
\end{figure}

Fig.~\ref{bimodal_stability} (middle and bottom) again shows numerical results of our agent-based model for a 30-regular network with $N=1000$ nodes and $r_0=0.67$ for the bimodal Dirac delta distribution for $\mu = 0$ (middle) and $\mu = 0.5$ (bottom) with the same procedure as Figs.~\ref{simstability} and \ref{het_stability} for each ordered pair $(\Delta,\epsilon)$. Fig.~\ref{bimodal_stability}  (top) shows an example of the bimodal Dirac delta distribution with mean $\mu$ and distance $\Delta$ between modes. In Fig.~\ref{bimodal_stability} (middle) we see that when the average opinion is zero, $\mu = 0$, increasing the distance between the two modes causes a shift from the disease-free state being stable when $\Delta = 0$ to a bistable region between the $+1$ and $-1$ endemic states when $\Delta  > 0$, for sufficiently small values of $\epsilon$. When the distribution is significantly biased to one side, $\mu = 0.5$, (Fig.~\ref{bimodal_stability} bottom) the difference between the two modes seems to have no effect. However, there is a transition between the $+1$ endemic and the disease-free state as $\epsilon$ is increased, mirroring what was found in the case of homogeneous opinions (cf. Fig.~\ref{stability}). 

The results of this section show that our model can result in rich behavior that is not captured by our mean field approximation when the initial distribution of opinions is highly heterogeneous. However, we leave an analysis of this case for future research.

\section{\label{reintroduction} External Recruitment of Spreaders}

Now we modify our model to allow for the external recruitment of spreaders. This could model a situation where disinformation is spread by the coordinated actions of malicious external agents. To model this we introduce an additional forcing term to our system which allows for external recruitment of contagion spreaders. Working in the framework of Eqs.~(\ref{SP})-(\ref{OP}), we modify them as 
\begin{eqnarray}
    \frac{dS_+}{d\tau}&=&-S_+\left[1-X+\epsilon \right]\notag \\ &+&r_0(1-S_+-S_-)S_+\left[1+X+\epsilon\right]\notag \\ &+& (1-S_+-S_-)f_+(t)\, ,\label{SPforced}\\
    \frac{dS_-}{d\tau}&=&-S_-\left[1+X+\epsilon\right]\notag \\ &+&r_0(1-S_+-S_-)S_-\left[1-X+\epsilon\right]\notag\\ &+&(1-S_+-S_-)f_-(t)\, ,\label{SNforced}\\
    \frac{dX}{d\tau}&=&K\left[(S_+-S_-)-(S_+ +S_-)X \right]\, ,\label{Xforced}
\end{eqnarray}
where $f_+(t)$ and $f_-(t)$ represent normalized rates of recruitment of spreaders for the $+1$ and $-1$ contagions, respectively. Since we are interested in how the spread of disinformation may affect the spread of the ``true" information, from this point on we will consider the $+1$ contagion as ``true" and the $-1$ contagion as ``false", recognizing that sometimes it is not possible to make such a clear distinction. In addition, for simplicity we will assume that only the ``false" contagion will have external recruitment of spreaders, meaning we will consider only the case where $f_+(t)=0$. For $f_-(t)$ we will consider only the case of constant forcing $f_-(t)=B$, which could represent a constant recruitment of ``false" information spreaders due to, for example, an unchanging social-bot network. With the addition of external forcing of the $-1$ contagion, disease-free behavior is no longer an equilibrium state of the system. Instead, with a constant forcing $f_-(t)=B$ there are now two steady-state equilibria. The first is of the form $(X,S_+,S_-)=(-1,0,S_-^*)$ where
\begin{eqnarray}
    S_-^*&=&\frac{-[B+\epsilon-r_0(2+\epsilon)]}{2r_0(2+\epsilon)} \notag\\
    &+& \frac{\sqrt{[B+\epsilon-r_0(2+\epsilon)]^2+4r_0(2+\epsilon)B}}{2r_0(2+\epsilon)}\, ,
\end{eqnarray}
which corresponds to the ``false" information becoming dominant in the system. The second is of the form $(X,S_+,S_-)=(X^*,S_2^*(X^*),S_1^*(X^*))$ where $X^*$, $S_2^*$, and $S_1^*$ are solutions to the non-linear algebraic equations
\begin{eqnarray}
    S_1^*(X^*)&=&\frac{B}{(1+X^*+\epsilon)\left(R^{+}_0(X^*) - R^{-}_0(X^*)\right)}\, ,\label{SN*}\\
    S_2^*(X^*)&=&1-\frac{1}{R^{+}_0(X^*)}-S_1^*(X^*)\, ,\label{SP*}\\
    X^*&=& 1 -\left[ \frac{2 R_0^{+}(X^*)}{R_0^{+}(X^*) - 1}\right]S_1^*(X^*)\, .
\end{eqnarray}
 This second equilibrium corresponds to the $+1$ contagion becoming dominant in the system while the $-1$ contagion remains sustained by a small fraction of the population, due to the constant external recruitment. We have found numerically that both of these solutions are stable. 

\begin{figure}
    \centering
    \includegraphics[width = \linewidth]{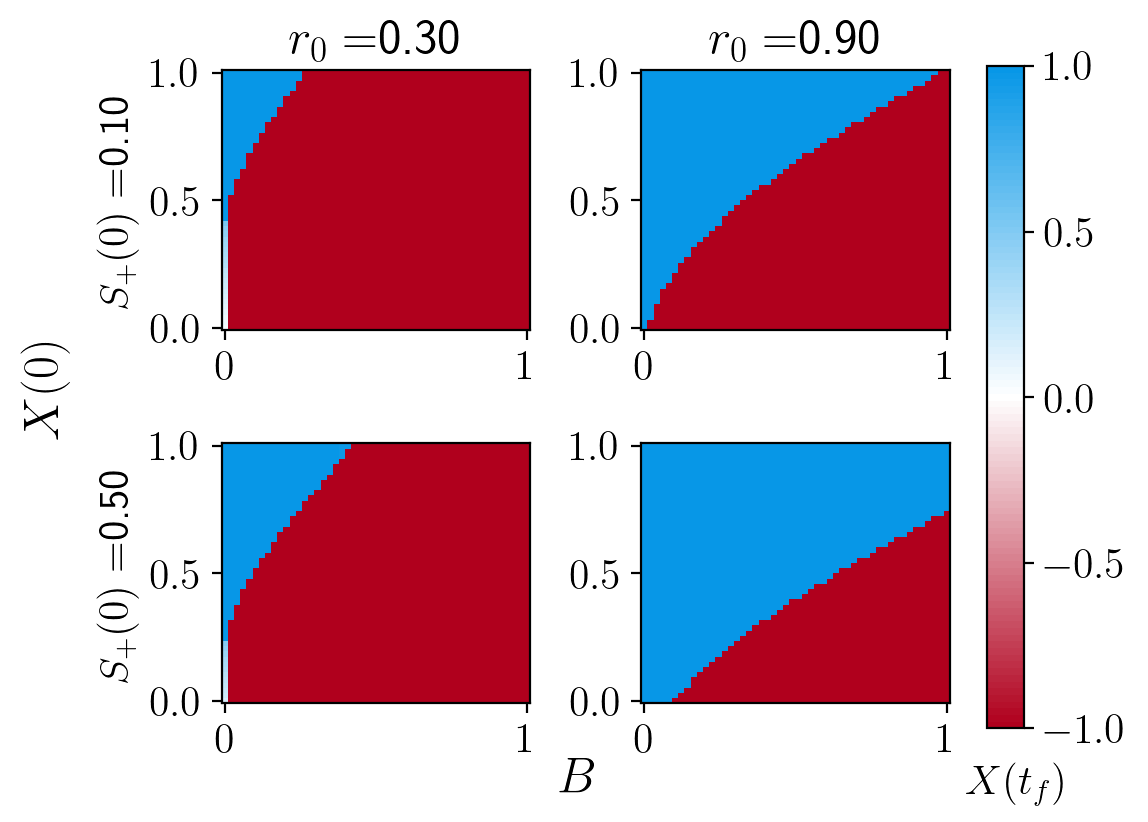}
    \caption{Average opinion of the population as a function of $X(0)$ and the constant forcing $f_-(t)=B$ after $t_f=3000$ time steps with parameters $\tau=0.1$, $K=0.25$, $\epsilon=0.01$, $r_0\in \{0.3,0.9\}$, $S_+(0)\in \{0.1,0.5\}$, and a time step $h=0.25$.}
    \label{heatmap_constant_forcing}
\end{figure}

Now we discuss how the forcing of the $-1$ contagion modifies the bias overturning behavior studied in Sec.~\ref{behavior}. In the absence of forcing, the bias overturning behavior is facilitated by a more infectious contagion: note how, in Fig.~\ref{IC_heatmap}, the red region (corresponding to the initial positive opinion being overturned) increases in size as $r_0$ increases. In contrast, a more infectious contagion suppresses bias overturning in the presence of constant external forcing. To illustrate this, Fig.~\ref{heatmap_constant_forcing} shows the average opinion after a long period of time ($t_f=3000$) against the forcing term $f_-(t)=B\in [0,1]$ and $X(0)\in [0,1]$ for $r_0=0.3,0.9$ and $S_+(0)=0.1,0.5$ with $S_-(0)=0$, obtained using the mean-field equations. Since $X(0)>0$ the system is initially biased towards the true contagion. In Fig.~\ref{heatmap_constant_forcing} we see that as $r_0$ increases the red region which, again, corresponds to the overturning behavior, becomes smaller. Therefore, in this case we see that for larger $r_0$ it is more difficult to overturn the initial bias. Similarly, as $S_+(0)$ increases a larger value of forcing is required to overturn the initial bias.

Although modeling an underlying social-bot network via external forcing terms is a limited approach, we observe within our model that information that spreads with lower values of $r_0$ is more susceptible to disinformation, as measured by the size of the basis of attraction of the $-1$ endemic state. 

\section{\label{discussion}Discussion}

We introduced a hybrid model of a pair of competing beliefs (interpreted as social contagions) coupled with an internal opinion describing the alignment of the individual's biases towards the two beliefs. Modeling cognitive biases, the internal opinion gets modified by infection attempts (modeling the illusory-truth effect) and modifies the infection probabilities (modeling confirmation bias). We found that this model results in an opinion dependent stability of the disease-free state (i.e., when the two beliefs are not being spread). In addition, we found that the incorporation of cognitive biases in the contagion process can lead to transient dynamical behaviors that are absent in simpler models of social contagions. These behaviors include the rebound, where one of the competing beliefs is revitalized after an initial decay, and the bias overturning, where the initial opinion of the population switches from one belief to the other. We found that bias overturning is promoted by stronger beliefs, as measured by the effective reproduction number $r_0$ in Eq.~(\ref{r0}); however, overturning an initial bias towards one belief when there is external recruitment of spreaders for the opposing belief is more difficult when the effective reproduction number is larger. While examining the role of heterogeneity in initial opinions we found that when the distribution of initial opinions is Gaussian the role of confirmation bias and the illusory truth effect decreases as the variance of the Gaussian increases. Furthermore, we observed that equally weighted bimodal distributions of initial opinions can result in bistability of the $\pm 1$ endemic behavior.

Despite the simplicity of our model, there are a number of results that can be related to the real-world spread of competing beliefs. First, we observe that the initial biasing of the system is highly important in the long-term behavior, both relating to the initial opinions of agents and the initial fraction spreading each belief. This suggests preemptive efforts to combat misinformation through prebunking type approaches are important in reducing the spread of false information. It also suggests that ignoring the early spread of false information may result in a system which is highly prone to belief in it and may be very resistant to recovery. Second, the formation of the rebound behavior may suggest that the effects of confirmation bias and the illusory-truth effect allow the creation of a type of ``reservoir" to the social contagion, similar to the existence of reservoirs in infectious contagions such as avian influenza and bovine tuberculosis \cite{simpson2002wild}, allowing the rebound of the belief into a population that initial rejected it. However, confirming such a feature in our model will require further study; particularly, closely examining the role of network and community structures as well as the spatial distribution of opinions within these structures. It should also be noted that the existence of such a rebound behavior and the relevant time scale is difficult to determine as it requires high quality time-dependent data on the spreading of beliefs in a social network. At this time we have been unable to find sufficient data to confirm or refute this feature of our model in real-world systems and would be an interesting and highly interdisciplinary extension of our study.

Our model is an idealized description of how two competing beliefs may spread in a regular network where individuals have cognitive biases such as confirmation bias and the illusory truth effect. There are many ways in which the model could be made more realistic. For example, our model doesn't account for potential interactions among more than two contagions, non-binary beliefs, realistic social network structure, fact-checking, or more detailed cognitive bias models. However, even within our highly simplified model there remain a number of potentially interesting questions. For example, in all analytical results, we assumed that identical initial opinions were assigned to all agents within the agent-based model. An extension of our analysis to the case of initial heterogeneous opinions is left for future work. Other examples include using more heterogeneous networks or real-world social networks instead of a $k$-regular network. 

\begin{acknowledgments}

CRS wants to acknowledge and thank Ekaterina Landgren, James Meiss, Mason Porter, Nancy Rodr\'{\i}guez, and Zachary Kilpatrick for useful comments and input. CRS also wants to acknowledge the use of the compleX Group Interactions (XGI) package for python and to thank the development team for such a useful and versatile toolkit \cite{Landry_XGI_2023}. JGR acknowledges support from NSF grant DMS-2205967. Lastly, we thank the referees for their useful comments.

\end{acknowledgments}

\appendix

\bibliography{Competing_Social_Contagions}

\begin{thebibliography}{41}%
\makeatletter
\providecommand \@ifxundefined [1]{%
 \@ifx{#1\undefined}
}%
\providecommand \@ifnum [1]{%
 \ifnum #1\expandafter \@firstoftwo
 \else \expandafter \@secondoftwo
 \fi
}%
\providecommand \@ifx [1]{%
 \ifx #1\expandafter \@firstoftwo
 \else \expandafter \@secondoftwo
 \fi
}%
\providecommand \natexlab [1]{#1}%
\providecommand \enquote  [1]{``#1''}%
\providecommand \bibnamefont  [1]{#1}%
\providecommand \bibfnamefont [1]{#1}%
\providecommand \citenamefont [1]{#1}%
\providecommand \href@noop [0]{\@secondoftwo}%
\providecommand \href [0]{\begingroup \@sanitize@url \@href}%
\providecommand \@href[1]{\@@startlink{#1}\@@href}%
\providecommand \@@href[1]{\endgroup#1\@@endlink}%
\providecommand \@sanitize@url [0]{\catcode `\\12\catcode `\$12\catcode
  `\&12\catcode `\#12\catcode `\^12\catcode `\_12\catcode `\%12\relax}%
\providecommand \@@startlink[1]{}%
\providecommand \@@endlink[0]{}%
\providecommand \url  [0]{\begingroup\@sanitize@url \@url }%
\providecommand \@url [1]{\endgroup\@href {#1}{\urlprefix }}%
\providecommand \urlprefix  [0]{URL }%
\providecommand \Eprint [0]{\href }%
\providecommand \doibase [0]{https://doi.org/}%
\providecommand \selectlanguage [0]{\@gobble}%
\providecommand \bibinfo  [0]{\@secondoftwo}%
\providecommand \bibfield  [0]{\@secondoftwo}%
\providecommand \translation [1]{[#1]}%
\providecommand \BibitemOpen [0]{}%
\providecommand \bibitemStop [0]{}%
\providecommand \bibitemNoStop [0]{.\EOS\space}%
\providecommand \EOS [0]{\spacefactor3000\relax}%
\providecommand \BibitemShut  [1]{\csname bibitem#1\endcsname}%
\let\auto@bib@innerbib\@empty
\bibitem [{\citenamefont {Edosomwan}\ \emph {et~al.}(2011)\citenamefont
  {Edosomwan}, \citenamefont {Prakasan}, \citenamefont {Kouame}, \citenamefont
  {Watson},\ and\ \citenamefont {Seymour}}]{edosomwan2011history}%
  \BibitemOpen
  \bibfield  {author} {\bibinfo {author} {\bibfnamefont {S.}~\bibnamefont
  {Edosomwan}}, \bibinfo {author} {\bibfnamefont {S.~K.}\ \bibnamefont
  {Prakasan}}, \bibinfo {author} {\bibfnamefont {D.}~\bibnamefont {Kouame}},
  \bibinfo {author} {\bibfnamefont {J.}~\bibnamefont {Watson}},\ and\ \bibinfo
  {author} {\bibfnamefont {T.}~\bibnamefont {Seymour}},\ }\bibfield  {title}
  {\bibinfo {title} {The history of social media and its impact on business},\
  }\href@noop {} {\bibfield  {journal} {\bibinfo  {journal} {Journal of Applied
  Management and entrepreneurship}\ }\textbf {\bibinfo {volume} {16}},\
  \bibinfo {pages} {79} (\bibinfo {year} {2011})}\BibitemShut {NoStop}%
\bibitem [{\citenamefont {Sajithra}\ and\ \citenamefont
  {Patil}(2013)}]{sajithra2013social}%
  \BibitemOpen
  \bibfield  {author} {\bibinfo {author} {\bibfnamefont {K.}~\bibnamefont
  {Sajithra}}\ and\ \bibinfo {author} {\bibfnamefont {R.}~\bibnamefont
  {Patil}},\ }\bibfield  {title} {\bibinfo {title} {Social media--history and
  components},\ }\href@noop {} {\bibfield  {journal} {\bibinfo  {journal}
  {Journal of Business and Management}\ }\textbf {\bibinfo {volume} {7}},\
  \bibinfo {pages} {69} (\bibinfo {year} {2013})}\BibitemShut {NoStop}%
\bibitem [{\citenamefont {Shearer}\ and\ \citenamefont
  {Matsa}(2018)}]{shearer2018news}%
  \BibitemOpen
  \bibfield  {author} {\bibinfo {author} {\bibfnamefont {E.}~\bibnamefont
  {Shearer}}\ and\ \bibinfo {author} {\bibfnamefont {K.}~\bibnamefont
  {Matsa}},\ }\href@noop {} {\emph {\bibinfo {title} {News use across social
  media platforms in 2018}}},\ \bibinfo {type} {Tech. Rep.}\ (\bibinfo
  {institution} {Pew Research Center},\ \bibinfo {year} {2018})\BibitemShut
  {NoStop}%
\bibitem [{\citenamefont {Chadwick}\ and\ \citenamefont
  {Vaccari}(2019)}]{chadwick2019news}%
  \BibitemOpen
  \bibfield  {author} {\bibinfo {author} {\bibfnamefont {A.}~\bibnamefont
  {Chadwick}}\ and\ \bibinfo {author} {\bibfnamefont {C.}~\bibnamefont
  {Vaccari}},\ }\href@noop {} {\emph {\bibinfo {title} {News sharing on UK
  social media: Misinformation, disinformation, and correction}}},\ \bibinfo
  {type} {Tech. Rep.}\ (\bibinfo  {institution} {Loughborough University},\
  \bibinfo {year} {2019})\BibitemShut {NoStop}%
\bibitem [{\citenamefont {Liedke}\ and\ \citenamefont
  {Wang}(2023)}]{PewResearch2023}%
  \BibitemOpen
  \bibfield  {author} {\bibinfo {author} {\bibnamefont {Liedke}}\ and\ \bibinfo
  {author} {\bibnamefont {Wang}},\ }\href@noop {} {\emph {\bibinfo {title}
  {Social Media and News Fact Sheet}}},\ \bibinfo {type} {Tech. Rep.}\
  (\bibinfo  {institution} {Pew Research Center},\ \bibinfo {year}
  {2023})\BibitemShut {NoStop}%
\bibitem [{\citenamefont {Allcott}\ and\ \citenamefont
  {Gentzkow}(2017)}]{allcott2017social}%
  \BibitemOpen
  \bibfield  {author} {\bibinfo {author} {\bibfnamefont {H.}~\bibnamefont
  {Allcott}}\ and\ \bibinfo {author} {\bibfnamefont {M.}~\bibnamefont
  {Gentzkow}},\ }\bibfield  {title} {\bibinfo {title} {Social media and fake
  news in the 2016 election},\ }\href@noop {} {\bibfield  {journal} {\bibinfo
  {journal} {Journal of economic perspectives}\ }\textbf {\bibinfo {volume}
  {31}},\ \bibinfo {pages} {211} (\bibinfo {year} {2017})}\BibitemShut
  {NoStop}%
\bibitem [{\citenamefont {Bakshy}\ \emph {et~al.}(2015)\citenamefont {Bakshy},
  \citenamefont {Messing},\ and\ \citenamefont {Adamic}}]{bakshy2015exposure}%
  \BibitemOpen
  \bibfield  {author} {\bibinfo {author} {\bibfnamefont {E.}~\bibnamefont
  {Bakshy}}, \bibinfo {author} {\bibfnamefont {S.}~\bibnamefont {Messing}},\
  and\ \bibinfo {author} {\bibfnamefont {L.~A.}\ \bibnamefont {Adamic}},\
  }\bibfield  {title} {\bibinfo {title} {Exposure to ideologically diverse news
  and opinion on facebook},\ }\href@noop {} {\bibfield  {journal} {\bibinfo
  {journal} {Science}\ }\textbf {\bibinfo {volume} {348}},\ \bibinfo {pages}
  {1130} (\bibinfo {year} {2015})}\BibitemShut {NoStop}%
\bibitem [{\citenamefont {Di~Domenico}\ \emph {et~al.}(2021)\citenamefont
  {Di~Domenico}, \citenamefont {Sit}, \citenamefont {Ishizaka},\ and\
  \citenamefont {Nunan}}]{di2021fake}%
  \BibitemOpen
  \bibfield  {author} {\bibinfo {author} {\bibfnamefont {G.}~\bibnamefont
  {Di~Domenico}}, \bibinfo {author} {\bibfnamefont {J.}~\bibnamefont {Sit}},
  \bibinfo {author} {\bibfnamefont {A.}~\bibnamefont {Ishizaka}},\ and\
  \bibinfo {author} {\bibfnamefont {D.}~\bibnamefont {Nunan}},\ }\bibfield
  {title} {\bibinfo {title} {Fake news, social media and marketing: A
  systematic review},\ }\href@noop {} {\bibfield  {journal} {\bibinfo
  {journal} {Journal of Business Research}\ }\textbf {\bibinfo {volume}
  {124}},\ \bibinfo {pages} {329} (\bibinfo {year} {2021})}\BibitemShut
  {NoStop}%
\bibitem [{bla(2023)}]{blair2010annual}%
  \BibitemOpen
  \href@noop {} {\emph {\bibinfo {title} {Annual Threat Assessment of the US
  Intelligence Community}}},\ \bibinfo {type} {Tech. Rep.}\ (\bibinfo
  {institution} {Office of the Director of National Intelligence Washington,
  DC},\ \bibinfo {year} {2023})\BibitemShut {NoStop}%
\bibitem [{\citenamefont {Gelfert}(2018)}]{gelfert2018fake}%
  \BibitemOpen
  \bibfield  {author} {\bibinfo {author} {\bibfnamefont {A.}~\bibnamefont
  {Gelfert}},\ }\bibfield  {title} {\bibinfo {title} {Fake news: A
  definition},\ }\href@noop {} {\bibfield  {journal} {\bibinfo  {journal}
  {Informal logic}\ }\textbf {\bibinfo {volume} {38}},\ \bibinfo {pages} {84}
  (\bibinfo {year} {2018})}\BibitemShut {NoStop}%
\bibitem [{\citenamefont {Berthon}\ and\ \citenamefont
  {Pitt}(2018)}]{berthon2018brands}%
  \BibitemOpen
  \bibfield  {author} {\bibinfo {author} {\bibfnamefont {P.~R.}\ \bibnamefont
  {Berthon}}\ and\ \bibinfo {author} {\bibfnamefont {L.~F.}\ \bibnamefont
  {Pitt}},\ }\bibfield  {title} {\bibinfo {title} {Brands, truthiness and
  post-fact: managing brands in a post-rational world},\ }\href@noop {}
  {\bibfield  {journal} {\bibinfo  {journal} {Journal of Macromarketing}\
  }\textbf {\bibinfo {volume} {38}},\ \bibinfo {pages} {218} (\bibinfo {year}
  {2018})}\BibitemShut {NoStop}%
\bibitem [{\citenamefont {Rini}(2017)}]{rini2017fake}%
  \BibitemOpen
  \bibfield  {author} {\bibinfo {author} {\bibfnamefont {R.}~\bibnamefont
  {Rini}},\ }\bibfield  {title} {\bibinfo {title} {Fake news and partisan
  epistemology},\ }\href@noop {} {\bibfield  {journal} {\bibinfo  {journal}
  {Kennedy Institute of Ethics Journal}\ }\textbf {\bibinfo {volume} {27}},\
  \bibinfo {pages} {E} (\bibinfo {year} {2017})}\BibitemShut {NoStop}%
\bibitem [{\citenamefont {Lazer}\ \emph {et~al.}(2018)\citenamefont {Lazer},
  \citenamefont {Baum}, \citenamefont {Benkler}, \citenamefont {Berinsky},
  \citenamefont {Greenhill}, \citenamefont {Menczer}, \citenamefont {Metzger},
  \citenamefont {Nyhan}, \citenamefont {Pennycook}, \citenamefont {Rothschild}
  \emph {et~al.}}]{lazer2018science}%
  \BibitemOpen
  \bibfield  {author} {\bibinfo {author} {\bibfnamefont {D.~M.}\ \bibnamefont
  {Lazer}}, \bibinfo {author} {\bibfnamefont {M.~A.}\ \bibnamefont {Baum}},
  \bibinfo {author} {\bibfnamefont {Y.}~\bibnamefont {Benkler}}, \bibinfo
  {author} {\bibfnamefont {A.~J.}\ \bibnamefont {Berinsky}}, \bibinfo {author}
  {\bibfnamefont {K.~M.}\ \bibnamefont {Greenhill}}, \bibinfo {author}
  {\bibfnamefont {F.}~\bibnamefont {Menczer}}, \bibinfo {author} {\bibfnamefont
  {M.~J.}\ \bibnamefont {Metzger}}, \bibinfo {author} {\bibfnamefont
  {B.}~\bibnamefont {Nyhan}}, \bibinfo {author} {\bibfnamefont
  {G.}~\bibnamefont {Pennycook}}, \bibinfo {author} {\bibfnamefont
  {D.}~\bibnamefont {Rothschild}}, \emph {et~al.},\ }\bibfield  {title}
  {\bibinfo {title} {The science of fake news},\ }\href@noop {} {\bibfield
  {journal} {\bibinfo  {journal} {Science}\ }\textbf {\bibinfo {volume}
  {359}},\ \bibinfo {pages} {1094} (\bibinfo {year} {2018})}\BibitemShut
  {NoStop}%
\bibitem [{\citenamefont {Pennycook}\ \emph {et~al.}(2018)\citenamefont
  {Pennycook}, \citenamefont {Cannon},\ and\ \citenamefont
  {Rand}}]{pennycook2018prior}%
  \BibitemOpen
  \bibfield  {author} {\bibinfo {author} {\bibfnamefont {G.}~\bibnamefont
  {Pennycook}}, \bibinfo {author} {\bibfnamefont {T.~D.}\ \bibnamefont
  {Cannon}},\ and\ \bibinfo {author} {\bibfnamefont {D.~G.}\ \bibnamefont
  {Rand}},\ }\bibfield  {title} {\bibinfo {title} {Prior exposure increases
  perceived accuracy of fake news.},\ }\href@noop {} {\bibfield  {journal}
  {\bibinfo  {journal} {Journal of experimental psychology: general}\ }\textbf
  {\bibinfo {volume} {147}},\ \bibinfo {pages} {1865} (\bibinfo {year}
  {2018})}\BibitemShut {NoStop}%
\bibitem [{\citenamefont {Bronstein}\ \emph {et~al.}(2019)\citenamefont
  {Bronstein}, \citenamefont {Pennycook}, \citenamefont {Bear}, \citenamefont
  {Rand},\ and\ \citenamefont {Cannon}}]{bronstein2019belief}%
  \BibitemOpen
  \bibfield  {author} {\bibinfo {author} {\bibfnamefont {M.~V.}\ \bibnamefont
  {Bronstein}}, \bibinfo {author} {\bibfnamefont {G.}~\bibnamefont
  {Pennycook}}, \bibinfo {author} {\bibfnamefont {A.}~\bibnamefont {Bear}},
  \bibinfo {author} {\bibfnamefont {D.~G.}\ \bibnamefont {Rand}},\ and\
  \bibinfo {author} {\bibfnamefont {T.~D.}\ \bibnamefont {Cannon}},\ }\bibfield
   {title} {\bibinfo {title} {Belief in fake news is associated with
  delusionality, dogmatism, religious fundamentalism, and reduced analytic
  thinking},\ }\href@noop {} {\bibfield  {journal} {\bibinfo  {journal}
  {Journal of applied research in memory and cognition}\ }\textbf {\bibinfo
  {volume} {8}},\ \bibinfo {pages} {108} (\bibinfo {year} {2019})}\BibitemShut
  {NoStop}%
\bibitem [{\citenamefont {McDermott}(2019)}]{mcdermott2019psychological}%
  \BibitemOpen
  \bibfield  {author} {\bibinfo {author} {\bibfnamefont {R.}~\bibnamefont
  {McDermott}},\ }\bibfield  {title} {\bibinfo {title} {Psychological
  underpinnings of post-truth in political beliefs},\ }\href@noop {} {\bibfield
   {journal} {\bibinfo  {journal} {PS: Political Science \& Politics}\ }\textbf
  {\bibinfo {volume} {52}},\ \bibinfo {pages} {218} (\bibinfo {year}
  {2019})}\BibitemShut {NoStop}%
\bibitem [{\citenamefont {Vicario}\ \emph {et~al.}(2019)\citenamefont
  {Vicario}, \citenamefont {Quattrociocchi}, \citenamefont {Scala},\ and\
  \citenamefont {Zollo}}]{vicario2019polarization}%
  \BibitemOpen
  \bibfield  {author} {\bibinfo {author} {\bibfnamefont {M.~D.}\ \bibnamefont
  {Vicario}}, \bibinfo {author} {\bibfnamefont {W.}~\bibnamefont
  {Quattrociocchi}}, \bibinfo {author} {\bibfnamefont {A.}~\bibnamefont
  {Scala}},\ and\ \bibinfo {author} {\bibfnamefont {F.}~\bibnamefont {Zollo}},\
  }\bibfield  {title} {\bibinfo {title} {Polarization and fake news: Early
  warning of potential misinformation targets},\ }\href@noop {} {\bibfield
  {journal} {\bibinfo  {journal} {ACM Transactions on the Web (TWEB)}\ }\textbf
  {\bibinfo {volume} {13}},\ \bibinfo {pages} {1} (\bibinfo {year}
  {2019})}\BibitemShut {NoStop}%
\bibitem [{\citenamefont {Tandoc~Jr}\ \emph {et~al.}(2018)\citenamefont
  {Tandoc~Jr}, \citenamefont {Ling}, \citenamefont {Westlund}, \citenamefont
  {Duffy}, \citenamefont {Goh},\ and\ \citenamefont
  {Zheng~Wei}}]{tandoc2018audiences}%
  \BibitemOpen
  \bibfield  {author} {\bibinfo {author} {\bibfnamefont {E.~C.}\ \bibnamefont
  {Tandoc~Jr}}, \bibinfo {author} {\bibfnamefont {R.}~\bibnamefont {Ling}},
  \bibinfo {author} {\bibfnamefont {O.}~\bibnamefont {Westlund}}, \bibinfo
  {author} {\bibfnamefont {A.}~\bibnamefont {Duffy}}, \bibinfo {author}
  {\bibfnamefont {D.}~\bibnamefont {Goh}},\ and\ \bibinfo {author}
  {\bibfnamefont {L.}~\bibnamefont {Zheng~Wei}},\ }\bibfield  {title} {\bibinfo
  {title} {Audiences’ acts of authentication in the age of fake news: A
  conceptual framework},\ }\href@noop {} {\bibfield  {journal} {\bibinfo
  {journal} {New media \& society}\ }\textbf {\bibinfo {volume} {20}},\
  \bibinfo {pages} {2745} (\bibinfo {year} {2018})}\BibitemShut {NoStop}%
\bibitem [{\citenamefont {Cheng}\ and\ \citenamefont
  {Chen}(2021)}]{cheng2021influence}%
  \BibitemOpen
  \bibfield  {author} {\bibinfo {author} {\bibfnamefont {Y.}~\bibnamefont
  {Cheng}}\ and\ \bibinfo {author} {\bibfnamefont {Z.~F.}\ \bibnamefont
  {Chen}},\ }\bibfield  {title} {\bibinfo {title} {The influence of presumed
  fake news influence: Examining public support for corporate corrective
  response, media literacy interventions, and governmental regulation},\ }in\
  \href@noop {} {\emph {\bibinfo {booktitle} {What IS News?}}}\ (\bibinfo
  {publisher} {Routledge},\ \bibinfo {year} {2021})\ pp.\ \bibinfo {pages}
  {103--127}\BibitemShut {NoStop}%
\bibitem [{\citenamefont {Chen}\ and\ \citenamefont
  {Cheng}(2020)}]{chen2020consumer}%
  \BibitemOpen
  \bibfield  {author} {\bibinfo {author} {\bibfnamefont {Z.~F.}\ \bibnamefont
  {Chen}}\ and\ \bibinfo {author} {\bibfnamefont {Y.}~\bibnamefont {Cheng}},\
  }\bibfield  {title} {\bibinfo {title} {Consumer response to fake news about
  brands on social media: the effects of self-efficacy, media trust, and
  persuasion knowledge on brand trust},\ }\href@noop {} {\bibfield  {journal}
  {\bibinfo  {journal} {Journal of Product \& Brand Management}\ }\textbf
  {\bibinfo {volume} {29}},\ \bibinfo {pages} {188} (\bibinfo {year}
  {2020})}\BibitemShut {NoStop}%
\bibitem [{\citenamefont {Vafeiadis}\ \emph {et~al.}(2020)\citenamefont
  {Vafeiadis}, \citenamefont {Bortree}, \citenamefont {Buckley}, \citenamefont
  {Diddi},\ and\ \citenamefont {Xiao}}]{vafeiadis2020refuting}%
  \BibitemOpen
  \bibfield  {author} {\bibinfo {author} {\bibfnamefont {M.}~\bibnamefont
  {Vafeiadis}}, \bibinfo {author} {\bibfnamefont {D.~S.}\ \bibnamefont
  {Bortree}}, \bibinfo {author} {\bibfnamefont {C.}~\bibnamefont {Buckley}},
  \bibinfo {author} {\bibfnamefont {P.}~\bibnamefont {Diddi}},\ and\ \bibinfo
  {author} {\bibfnamefont {A.}~\bibnamefont {Xiao}},\ }\bibfield  {title}
  {\bibinfo {title} {Refuting fake news on social media: nonprofits, crisis
  response strategies and issue involvement},\ }\href@noop {} {\bibfield
  {journal} {\bibinfo  {journal} {Journal of Product \& Brand Management}\
  }\textbf {\bibinfo {volume} {29}},\ \bibinfo {pages} {209} (\bibinfo {year}
  {2020})}\BibitemShut {NoStop}%
\bibitem [{\citenamefont {Franceschi}\ and\ \citenamefont
  {Pareschi}(2022)}]{franceschi2022spreading}%
  \BibitemOpen
  \bibfield  {author} {\bibinfo {author} {\bibfnamefont {J.}~\bibnamefont
  {Franceschi}}\ and\ \bibinfo {author} {\bibfnamefont {L.}~\bibnamefont
  {Pareschi}},\ }\bibfield  {title} {\bibinfo {title} {Spreading of fake news,
  competence and learning: kinetic modelling and numerical approximation},\
  }\href@noop {} {\bibfield  {journal} {\bibinfo  {journal} {Philosophical
  Transactions of the Royal Society A}\ }\textbf {\bibinfo {volume} {380}},\
  \bibinfo {pages} {20210159} (\bibinfo {year} {2022})}\BibitemShut {NoStop}%
\bibitem [{\citenamefont {Murayama}\ \emph {et~al.}(2021)\citenamefont
  {Murayama}, \citenamefont {Wakamiya}, \citenamefont {Aramaki},\ and\
  \citenamefont {Kobayashi}}]{murayama2021modeling}%
  \BibitemOpen
  \bibfield  {author} {\bibinfo {author} {\bibfnamefont {T.}~\bibnamefont
  {Murayama}}, \bibinfo {author} {\bibfnamefont {S.}~\bibnamefont {Wakamiya}},
  \bibinfo {author} {\bibfnamefont {E.}~\bibnamefont {Aramaki}},\ and\ \bibinfo
  {author} {\bibfnamefont {R.}~\bibnamefont {Kobayashi}},\ }\bibfield  {title}
  {\bibinfo {title} {Modeling the spread of fake news on twitter},\ }\href@noop
  {} {\bibfield  {journal} {\bibinfo  {journal} {Plos one}\ }\textbf {\bibinfo
  {volume} {16}},\ \bibinfo {pages} {e0250419} (\bibinfo {year}
  {2021})}\BibitemShut {NoStop}%
\bibitem [{\citenamefont {Rabb}\ \emph {et~al.}(2022)\citenamefont {Rabb},
  \citenamefont {Cowen}, \citenamefont {de~Ruiter},\ and\ \citenamefont
  {Scheutz}}]{rabb2022cognitive}%
  \BibitemOpen
  \bibfield  {author} {\bibinfo {author} {\bibfnamefont {N.}~\bibnamefont
  {Rabb}}, \bibinfo {author} {\bibfnamefont {L.}~\bibnamefont {Cowen}},
  \bibinfo {author} {\bibfnamefont {J.~P.}\ \bibnamefont {de~Ruiter}},\ and\
  \bibinfo {author} {\bibfnamefont {M.}~\bibnamefont {Scheutz}},\ }\bibfield
  {title} {\bibinfo {title} {Cognitive cascades: How to model (and potentially
  counter) the spread of fake news},\ }\href@noop {} {\bibfield  {journal}
  {\bibinfo  {journal} {Plos one}\ }\textbf {\bibinfo {volume} {17}},\ \bibinfo
  {pages} {e0261811} (\bibinfo {year} {2022})}\BibitemShut {NoStop}%
\bibitem [{\citenamefont {Tambuscio}\ \emph {et~al.}(2015)\citenamefont
  {Tambuscio}, \citenamefont {Ruffo}, \citenamefont {Flammini},\ and\
  \citenamefont {Menczer}}]{tambuscio2015fact}%
  \BibitemOpen
  \bibfield  {author} {\bibinfo {author} {\bibfnamefont {M.}~\bibnamefont
  {Tambuscio}}, \bibinfo {author} {\bibfnamefont {G.}~\bibnamefont {Ruffo}},
  \bibinfo {author} {\bibfnamefont {A.}~\bibnamefont {Flammini}},\ and\
  \bibinfo {author} {\bibfnamefont {F.}~\bibnamefont {Menczer}},\ }\bibfield
  {title} {\bibinfo {title} {Fact-checking effect on viral hoaxes: A model of
  misinformation spread in social networks},\ }in\ \href@noop {} {\emph
  {\bibinfo {booktitle} {Proceedings of the 24th international conference on
  World Wide Web}}}\ (\bibinfo {year} {2015})\ pp.\ \bibinfo {pages}
  {977--982}\BibitemShut {NoStop}%
\bibitem [{\citenamefont {Budak}\ \emph {et~al.}(2011)\citenamefont {Budak},
  \citenamefont {Agrawal},\ and\ \citenamefont
  {El~Abbadi}}]{budak2011limiting}%
  \BibitemOpen
  \bibfield  {author} {\bibinfo {author} {\bibfnamefont {C.}~\bibnamefont
  {Budak}}, \bibinfo {author} {\bibfnamefont {D.}~\bibnamefont {Agrawal}},\
  and\ \bibinfo {author} {\bibfnamefont {A.}~\bibnamefont {El~Abbadi}},\
  }\bibfield  {title} {\bibinfo {title} {Limiting the spread of misinformation
  in social networks},\ }in\ \href@noop {} {\emph {\bibinfo {booktitle}
  {Proceedings of the 20th international conference on World wide web}}}\
  (\bibinfo {year} {2011})\ pp.\ \bibinfo {pages} {665--674}\BibitemShut
  {NoStop}%
\bibitem [{\citenamefont {Zareie}\ and\ \citenamefont
  {Sakellariou}(2021)}]{zareie2021minimizing}%
  \BibitemOpen
  \bibfield  {author} {\bibinfo {author} {\bibfnamefont {A.}~\bibnamefont
  {Zareie}}\ and\ \bibinfo {author} {\bibfnamefont {R.}~\bibnamefont
  {Sakellariou}},\ }\bibfield  {title} {\bibinfo {title} {Minimizing the spread
  of misinformation in online social networks: A survey},\ }\href@noop {}
  {\bibfield  {journal} {\bibinfo  {journal} {Journal of Network and Computer
  Applications}\ }\textbf {\bibinfo {volume} {186}},\ \bibinfo {pages} {103094}
  (\bibinfo {year} {2021})}\BibitemShut {NoStop}%
\bibitem [{\citenamefont {Van Der~Linden}(2022)}]{van2022misinformation}%
  \BibitemOpen
  \bibfield  {author} {\bibinfo {author} {\bibfnamefont {S.}~\bibnamefont {Van
  Der~Linden}},\ }\bibfield  {title} {\bibinfo {title} {Misinformation:
  susceptibility, spread, and interventions to immunize the public},\
  }\href@noop {} {\bibfield  {journal} {\bibinfo  {journal} {Nature Medicine}\
  }\textbf {\bibinfo {volume} {28}},\ \bibinfo {pages} {460} (\bibinfo {year}
  {2022})}\BibitemShut {NoStop}%
\bibitem [{\citenamefont {Cook}\ \emph {et~al.}(2015)\citenamefont {Cook},
  \citenamefont {Ecker},\ and\ \citenamefont
  {Lewandowsky}}]{cook2015misinformation}%
  \BibitemOpen
  \bibfield  {author} {\bibinfo {author} {\bibfnamefont {J.}~\bibnamefont
  {Cook}}, \bibinfo {author} {\bibfnamefont {U.}~\bibnamefont {Ecker}},\ and\
  \bibinfo {author} {\bibfnamefont {S.}~\bibnamefont {Lewandowsky}},\
  }\bibfield  {title} {\bibinfo {title} {Misinformation and how to correct
  it},\ }\href@noop {} {\bibfield  {journal} {\bibinfo  {journal} {Emerging
  trends in the social and behavioral sciences: An interdisciplinary,
  searchable, and linkable resource}\ ,\ \bibinfo {pages} {1}} (\bibinfo {year}
  {2015})}\BibitemShut {NoStop}%
\bibitem [{\citenamefont {BIAS}(2015)}]{bias2015evolution}%
  \BibitemOpen
  \bibfield  {author} {\bibinfo {author} {\bibfnamefont {F.}~\bibnamefont
  {BIAS}},\ }\bibfield  {title} {\bibinfo {title} {The evolution of cognitive
  bias},\ }\href@noop {} {\bibfield  {journal} {\bibinfo  {journal} {The
  Handbook of Evolutionary Psychology}\ }\textbf {\bibinfo {volume} {2}}
  (\bibinfo {year} {2015})}\BibitemShut {NoStop}%
\bibitem [{\citenamefont {Hart}\ \emph {et~al.}(2009)\citenamefont {Hart},
  \citenamefont {Albarrac{\'\i}n}, \citenamefont {Eagly}, \citenamefont
  {Brechan}, \citenamefont {Lindberg},\ and\ \citenamefont
  {Merrill}}]{hart2009feeling}%
  \BibitemOpen
  \bibfield  {author} {\bibinfo {author} {\bibfnamefont {W.}~\bibnamefont
  {Hart}}, \bibinfo {author} {\bibfnamefont {D.}~\bibnamefont
  {Albarrac{\'\i}n}}, \bibinfo {author} {\bibfnamefont {A.~H.}\ \bibnamefont
  {Eagly}}, \bibinfo {author} {\bibfnamefont {I.}~\bibnamefont {Brechan}},
  \bibinfo {author} {\bibfnamefont {M.~J.}\ \bibnamefont {Lindberg}},\ and\
  \bibinfo {author} {\bibfnamefont {L.}~\bibnamefont {Merrill}},\ }\bibfield
  {title} {\bibinfo {title} {Feeling validated versus being correct: a
  meta-analysis of selective exposure to information.},\ }\href@noop {}
  {\bibfield  {journal} {\bibinfo  {journal} {Psychological bulletin}\ }\textbf
  {\bibinfo {volume} {135}},\ \bibinfo {pages} {555} (\bibinfo {year}
  {2009})}\BibitemShut {NoStop}%
\bibitem [{\citenamefont {Hasher}\ \emph {et~al.}(1977)\citenamefont {Hasher},
  \citenamefont {Goldstein},\ and\ \citenamefont
  {Toppino}}]{hasher1977frequency}%
  \BibitemOpen
  \bibfield  {author} {\bibinfo {author} {\bibfnamefont {L.}~\bibnamefont
  {Hasher}}, \bibinfo {author} {\bibfnamefont {D.}~\bibnamefont {Goldstein}},\
  and\ \bibinfo {author} {\bibfnamefont {T.}~\bibnamefont {Toppino}},\
  }\bibfield  {title} {\bibinfo {title} {Frequency and the conference of
  referential validity},\ }\href@noop {} {\bibfield  {journal} {\bibinfo
  {journal} {Journal of verbal learning and verbal behavior}\ }\textbf
  {\bibinfo {volume} {16}},\ \bibinfo {pages} {107} (\bibinfo {year}
  {1977})}\BibitemShut {NoStop}%
\bibitem [{\citenamefont {Weng}\ \emph {et~al.}(2024)\citenamefont {Weng},
  \citenamefont {Johnson},\ and\ \citenamefont {Kreuter}}]{weng2024repeated}%
  \BibitemOpen
  \bibfield  {author} {\bibinfo {author} {\bibfnamefont {O.}~\bibnamefont
  {Weng}}, \bibinfo {author} {\bibfnamefont {K.~J.}\ \bibnamefont {Johnson}},\
  and\ \bibinfo {author} {\bibfnamefont {M.~W.}\ \bibnamefont {Kreuter}},\
  }\bibfield  {title} {\bibinfo {title} {Repeated exposure to covid-19
  misinformation: A longitudinal analysis of prevalence and predictors in a
  community sample},\ }\href@noop {} {\bibfield  {journal} {\bibinfo  {journal}
  {Journal of Public Health Management and Practice}\ }\textbf {\bibinfo
  {volume} {30}},\ \bibinfo {pages} {E211} (\bibinfo {year}
  {2024})}\BibitemShut {NoStop}%
\bibitem [{\citenamefont {Zhuang}\ \emph {et~al.}(2017)\citenamefont {Zhuang},
  \citenamefont {Chen},\ and\ \citenamefont {Li}}]{zhuang2017modeling}%
  \BibitemOpen
  \bibfield  {author} {\bibinfo {author} {\bibfnamefont {Y.-B.}\ \bibnamefont
  {Zhuang}}, \bibinfo {author} {\bibfnamefont {J.}~\bibnamefont {Chen}},\ and\
  \bibinfo {author} {\bibfnamefont {Z.-h.}\ \bibnamefont {Li}},\ }\bibfield
  {title} {\bibinfo {title} {Modeling the cooperative and competitive
  contagions in online social networks},\ }\href@noop {} {\bibfield  {journal}
  {\bibinfo  {journal} {Physica A: Statistical Mechanics and its Applications}\
  }\textbf {\bibinfo {volume} {484}},\ \bibinfo {pages} {141} (\bibinfo {year}
  {2017})}\BibitemShut {NoStop}%
\bibitem [{\citenamefont {Myers}\ and\ \citenamefont
  {Leskovec}(2012)}]{myers2012clash}%
  \BibitemOpen
  \bibfield  {author} {\bibinfo {author} {\bibfnamefont {S.~A.}\ \bibnamefont
  {Myers}}\ and\ \bibinfo {author} {\bibfnamefont {J.}~\bibnamefont
  {Leskovec}},\ }\bibfield  {title} {\bibinfo {title} {Clash of the contagions:
  Cooperation and competition in information diffusion},\ }in\ \href@noop {}
  {\emph {\bibinfo {booktitle} {2012 IEEE 12th international conference on data
  mining}}}\ (\bibinfo {organization} {IEEE},\ \bibinfo {year} {2012})\ pp.\
  \bibinfo {pages} {539--548}\BibitemShut {NoStop}%
\bibitem [{\citenamefont {Fu}\ \emph {et~al.}(2017)\citenamefont {Fu},
  \citenamefont {Christakis},\ and\ \citenamefont {Fowler}}]{fu2017dueling}%
  \BibitemOpen
  \bibfield  {author} {\bibinfo {author} {\bibfnamefont {F.}~\bibnamefont
  {Fu}}, \bibinfo {author} {\bibfnamefont {N.~A.}\ \bibnamefont {Christakis}},\
  and\ \bibinfo {author} {\bibfnamefont {J.~H.}\ \bibnamefont {Fowler}},\
  }\bibfield  {title} {\bibinfo {title} {Dueling biological and social
  contagions},\ }\href@noop {} {\bibfield  {journal} {\bibinfo  {journal}
  {Scientific reports}\ }\textbf {\bibinfo {volume} {7}},\ \bibinfo {pages}
  {43634} (\bibinfo {year} {2017})}\BibitemShut {NoStop}%
\bibitem [{\citenamefont {Kaligotla}\ \emph {et~al.}(2015)\citenamefont
  {Kaligotla}, \citenamefont {Yücesan},\ and\ \citenamefont
  {Chick}}]{kaligotla2015}%
  \BibitemOpen
  \bibfield  {author} {\bibinfo {author} {\bibfnamefont {C.}~\bibnamefont
  {Kaligotla}}, \bibinfo {author} {\bibfnamefont {E.}~\bibnamefont
  {Yücesan}},\ and\ \bibinfo {author} {\bibfnamefont {S.~E.}\ \bibnamefont
  {Chick}},\ }\bibfield  {title} {\bibinfo {title} {An agent based model of
  spread of competing rumors through online interactions on social media},\
  }in\ \href {https://doi.org/10.1109/WSC.2015.7408553} {\emph {\bibinfo
  {booktitle} {2015 Winter Simulation Conference (WSC)}}}\ (\bibinfo {year}
  {2015})\ pp.\ \bibinfo {pages} {3985--3996}\BibitemShut {NoStop}%
\bibitem [{\citenamefont {Kiss}\ \emph {et~al.}(2017)\citenamefont {Kiss},
  \citenamefont {Miller},\ and\ \citenamefont {Simon}}]{istvan2017mathematics}%
  \BibitemOpen
  \bibfield  {author} {\bibinfo {author} {\bibfnamefont {I.~Z.}\ \bibnamefont
  {Kiss}}, \bibinfo {author} {\bibfnamefont {J.~S.}\ \bibnamefont {Miller}},\
  and\ \bibinfo {author} {\bibfnamefont {P.}~\bibnamefont {Simon}},\
  }\href@noop {} {\emph {\bibinfo {title} {Mathematics of Epidemics on
  Networks: From Exact to Approximate Models}}}\ (\bibinfo  {publisher}
  {Springer International Publishing},\ \bibinfo {year} {2017})\BibitemShut
  {NoStop}%
\bibitem [{\citenamefont {Landry}\ \emph {et~al.}(2023)\citenamefont {Landry},
  \citenamefont {Lucas}, \citenamefont {Iacopini}, \citenamefont {Petri},
  \citenamefont {Schwarze}, \citenamefont {Patania},\ and\ \citenamefont
  {Torres}}]{Landry_XGI_2023}%
  \BibitemOpen
  \bibfield  {author} {\bibinfo {author} {\bibfnamefont {N.~W.}\ \bibnamefont
  {Landry}}, \bibinfo {author} {\bibfnamefont {M.}~\bibnamefont {Lucas}},
  \bibinfo {author} {\bibfnamefont {I.}~\bibnamefont {Iacopini}}, \bibinfo
  {author} {\bibfnamefont {G.}~\bibnamefont {Petri}}, \bibinfo {author}
  {\bibfnamefont {A.}~\bibnamefont {Schwarze}}, \bibinfo {author}
  {\bibfnamefont {A.}~\bibnamefont {Patania}},\ and\ \bibinfo {author}
  {\bibfnamefont {L.}~\bibnamefont {Torres}},\ }\bibfield  {title} {\bibinfo
  {title} {{XGI: A Python package for higher-order interaction networks}},\
  }\href {https://doi.org/10.21105/joss.05162} {\bibfield  {journal} {\bibinfo
  {journal} {Journal of Open Source Software}\ }\textbf {\bibinfo {volume}
  {8}},\ \bibinfo {pages} {5162} (\bibinfo {year} {2023})}\BibitemShut
  {NoStop}%
\bibitem [{\citenamefont {Lewandowsky}\ and\ \citenamefont {Van
  Der~Linden}(2021)}]{lewandowsky2021countering}%
  \BibitemOpen
  \bibfield  {author} {\bibinfo {author} {\bibfnamefont {S.}~\bibnamefont
  {Lewandowsky}}\ and\ \bibinfo {author} {\bibfnamefont {S.}~\bibnamefont {Van
  Der~Linden}},\ }\bibfield  {title} {\bibinfo {title} {Countering
  misinformation and fake news through inoculation and prebunking},\
  }\href@noop {} {\bibfield  {journal} {\bibinfo  {journal} {European Review of
  Social Psychology}\ }\textbf {\bibinfo {volume} {32}},\ \bibinfo {pages}
  {348} (\bibinfo {year} {2021})}\BibitemShut {NoStop}%
\bibitem [{\citenamefont {Simpson}(2002)}]{simpson2002wild}%
  \BibitemOpen
  \bibfield  {author} {\bibinfo {author} {\bibfnamefont {V.}~\bibnamefont
  {Simpson}},\ }\bibfield  {title} {\bibinfo {title} {Wild animals as
  reservoirs of infectious diseases in the uk},\ }\href@noop {} {\bibfield
  {journal} {\bibinfo  {journal} {The Veterinary Journal}\ }\textbf {\bibinfo
  {volume} {163}},\ \bibinfo {pages} {128} (\bibinfo {year}
  {2002})}\BibitemShut {NoStop}%
\end{thebibliography}%

\end{document}